%% file: acm-authordraft.tex
\documentclass[sigconf]{acmart}
\usepackage{tabulary}
\usepackage{enumitem}
\usepackage[]{collab}

\collabAuthor{zb}{teal}{Zhuangbin Chen}

\AtBeginDocument{%
  \providecommand\BibTeX{{%
    \normalfont B\kern-0.5em{\scshape i\kern-0.25em b}\kern-0.8em\TeX}}}

\setcopyright{acmcopyright}
\copyrightyear{2021}
\acmYear{2021}
\acmDOI{10.1145/1122445.1122456}

\acmConference[Woodstock '18]{Woodstock '18: ACM Symposium on Neural Gaze Detection}{June 03--05, 2018}{Woodstock, NY}
\acmBooktitle{Woodstock '18: ACM Symposium on Neural Gaze Detection, June 03--05, 2018, Woodstock, NY}
\acmPrice{15.00}
\acmISBN{978-1-4503-XXXX-X/18/06}
\begin{document}

\title{Experience Report: Deep Learning-based System Log Analysis for Anomaly Detection}

\author{Zhuangbin Chen}
\affiliation{%
  \institution{The Chinese University of Hong Kong}
  \city{Hong Kong}
  \country{China}}

\author{Jinyang Liu}
\author{Wenwei Gu}
\affiliation{%
  \institution{The Chinese University of Hong Kong}
  \city{Hong Kong}
  \country{China}}


\author{Yuxin Su}
\authornote{Corresponding author: suyx35@mail.sysu.edu.cn.}
\affiliation{
  \institution{Sun Yat-sen University}
  \city{Zhuhai}
  \country{China}}

\author{Jieming Zhu}
\affiliation{
  \institution{Huawei Noah’s Ark Lab}
  \city{Shenzhen}
  \country{China}}

\author{Yongqiang Yang}
\affiliation{
  \institution{Huawei Cloud BU}
  \city{Beijing}
  \country{China}}

\author{Michael R. Lyu}
\affiliation{
  \institution{The Chinese University of Hong Kong}
  \city{Hong Kong}
  \country{China}}



\begin{abstract}

Logs have been an imperative resource to ensure the reliability and continuity of many software systems, especially large-scale distributed systems. They faithfully record runtime information to facilitate system troubleshooting and behavior understanding. Due to the large scale and complexity of modern software systems, the volume of logs has reached an unprecedented level. Consequently, for log-based anomaly detection, conventional manual inspection methods or even traditional machine learning-based methods become impractical, which serve as a catalyst for the rapid development of deep learning-based solutions. However, there is currently a lack of rigorous comparison among the representative log-based anomaly detectors that resort to neural networks. Moreover, the re-implementation process demands non-trivial efforts, and bias can be easily introduced. To better understand the characteristics of different anomaly detectors, in this paper, we provide a comprehensive review and evaluation of five popular neural networks used by six state-of-the-art methods. Particularly, four of the selected methods are unsupervised, and the remaining two are supervised. These methods are evaluated with two publicly available log datasets, which contain nearly 16 million log messages and 0.4 million anomaly instances in total. We believe our work can serve as a basis in this field and contribute to future academic research and industrial applications.

\end{abstract}

\maketitle

\input{secs/intro}
\input{secs/overview}
\input{secs/deep_loglizer}
\input{secs/eval}

\input{secs/industry}
\input{secs/related_work}
\input{secs/conclusion}


\bibliographystyle{ACM-Reference-Format}
\bibliography{bibliography}


\end{document}

%% file: secs/intro.tex
\section{Introduction}
\label{sec:introduction}

Recent decades have witnessed an increasing prevalence of software systems providing a variety of services in our daily lives (e.g., search engines, social media, translation applications). Different from traditional on-premises software, modern software, e.g., online services, often serves hundreds of millions of customers worldwide with a goal of 24x7 availability~\cite{chen2020towards}. With such an unprecedented scale and complexity, how service failures and performance degradation are managed becomes a core competence on the market. Logs faithfully reflect the runtime status of a software system, which are of great importance for the monitoring, administering, and troubleshooting of a system. Therefore, log-based anomaly detection, which aims to uncover abnormal system behaviors, has become an essential means to ensure system reliability and service quality.

For traditional on-premise software systems, engineers usually perform simple keyword searches (such as ``failed'', ``exception'', and ``error'') or rule matching~\cite{hansen1993automated,prewett2003analyzing} to locate suspicious logs that might be associated with system problems. Due to the ever-increasing volume, variety, and velocity of logs produced by modern software systems, such manual approaches fall short for being labor-intensive and error-prone. Thus, many studies resort to statistical and traditional machine learning (ML) algorithms to incorporate more automation into this process. Exemplary algorithms are principal component analysis (PCA)~\cite{xu2009detecting}, log clustering~\cite{lin2016log}, etc.
Although these methods have made remarkable performance gains, they still face the following limitations in terms of practical deployments:

\begin{itemize}[noitemsep,topsep=0pt]
    \item \textit{Insufficient interpretability}. For log anomaly detection, interpretable results are critical for admins and analysts to trust and act on the automated analysis, e.g., which logs are important or which system components are problematic. However, many traditional methods only make a simple prediction for input with no further details. Engineers need to conduct a manual investigation for fault localization, which, in large-scale systems, is like finding a needle in a haystack.
    
    \item \textit{Weak adaptability}. During feature extraction, these methods often require the set of distinct log events to be known beforehand~\cite{zhang2019robust}. However, as modern systems continuously undergo feature addition and system upgrade, unseen log events could constantly emerge. To embrace the new log events, some models need to be retrained from scratch.
    
    \item \textit{Handcrafted features}. As an essential part of traditional ML workflow, many ML-based methods, e.g.,~\cite{zhou2019latent,liang2007failure}, require tailored features. Due to the variety of different systems, some of the selected features might not always be applicable, and other critical ones could be missing. Feature engineering is time-consuming and demands human domain knowledge.
    
\end{itemize}

Due to the exceptional ability in modeling complex relationships, deep learning (DL) has produced results comparable to and in some areas surpassing human expert performance. It often adopts a multiple-layer architecture called neural networks to progressively extract features from inputs with different layers dealing with different levels of feature abstraction. Typical architectures include recurrent neural networks (RNNs), convolutional neural networks (CNNs), autoencoder, etc. They have been widely applied to various fields, including computer vision, neural language processing, etc. In recent years, there has been an explosion of interest in applying neural networks to log-based anomaly detection. For example, Du et al.~\cite{du2017deeplog} employed long short-term memory (LSTM) networks for this purpose. On top of their work, Zhang et al.~\cite{zhang2019robust} and Meng et al.~\cite{meng2019loganomaly} further considered the semantic information of logs to improve the model's adaptability to unprecedented logs.

Given such fruitful achievements in the literature, we, however, observe a gap between academic research and industrial practices. One important reason is that site reliability engineers may not have fully realized the advances of DL techniques in log-based anomaly detection~\cite{dang2019aiops}. Thus, they are not aware of the existence of some state-of-the-art anomaly detection methods. This issue is further compounded by the fact that engineers may not have enough ML/data science background and skills.
As a result, it would be cumbersome for them to search through the literature and select the most appropriate method(s) for the problems at hand. Another important reason is that, to the best of our knowledge, there is currently no open-source toolkit available that applies DL techniques for log-based anomaly detection. Thus, if the code of the original paper is not open-source (which is not uncommon), engineers need to re-implement the model from scratch. In this process, bias and errors can be easily introduced because 1) the papers may not provide enough implementation details (e.g., parameter settings), and 2) engineers may lack experience developing DL models with relevant frameworks such as PyTorch~\cite{pytorch} and TensorFlow~\cite{tensorflow}.

He et al.~\cite{he2016experience} have conducted an important comparative study in this area, covering only traditional ML-based methods. Compared to them, DL-based methods possess the following merits: 1) more interpretable results, which are vital for engineers and analysts to take remediation actions, 2) better generalization ability to unseen logs which constantly appear in modern software systems, and 3) automated feature engineering which requires little human intervention. These merits render the necessity of a complementary study of the DL-based solutions. In this paper, we conduct a comprehensive review and evaluation of five representative neural networks used by six log anomaly detection methods. To facilitate reuse, we also release an open-source toolkit\footnote{\url{https://github.com/logpai/deep-loglizer}} containing the studied models. We believe researchers and practitioners can benefit from our work in the following two aspects: 1) they can quickly understand the characteristics of popular DL-based anomaly detectors and the differences with their traditional ML-based counterparts, and 2) they can save enormous efforts on re-implementations and focus on further customization or improvement.

The log anomaly detectors selected in this work include four unsupervised methods (i.e., two LSTMs~\cite{du2017deeplog,meng2019loganomaly}, the Transformer~\cite{nedelkoski2020self}, and Autoencoder~\cite{farzad2020unsupervised}) and two supervised methods (i.e., CNN~\cite{lu2018detecting} and attentional BiLSTM~\cite{zhang2019robust}). As labels are often unobtainable in real-world scenarios~\cite{chen2020towards}, unsupervised methods are more favored in the literature. When a system runs in a healthy state, the generated logs often exhibit stable and normal patterns. An abnormal instance usually manifests itself as an outlier that significantly deviates from such patterns. Based on this observation, unsupervised methods try to model logs' normal patterns and measure the deviation for each data instance. On the other hand, supervised methods directly learn the features that can best discriminate normal and abnormal cases based on the labels. All selected methods are evaluated on two widely-used log datasets that are publicly available, i.e., HDFS and BGL, containing nearly 16 million log messages and 0.4 million anomaly instances in total. The evaluation results are reported in \textit{accuracy}, \textit{robustness}, and \textit{efficiency}. We believe our work can prompt industrial applications of more recent log-based anomaly detection studies and provide guidelines for future research.

To sum up, this work makes the following major contributions:

\begin{itemize}
    \item We provide a comprehensive review of five representative neural networks used by six state-of-the-art DL-based log anomaly detectors.
    
    \item We release an open-source toolkit containing the studied methods to allow easy reuse for the community.
    
    \item We conduct a systematic evaluation that benchmarks the effectiveness and efficiency of the selected models and compare them with their traditional ML-based counterparts.
\end{itemize}

The remainder of this paper is organized as follows. Section~\ref{sec:overview} provides an overview about the process of log-based anomaly detection. Section~\ref{sec:deep_loglizer} summarizes the problem formulation of log anomaly detection and reviews six representative methods leveraging neural networks. Section~\ref{sec:evaluation} presents the experiments and experimental results. Section~\ref{sec:industry} shares our industrial practices. Section~\ref{sec:related_work} discusses some related work. Finally, Section~\ref{sec:conclusion} concludes this work.

%% file: secs/overview.tex
\section{Log Anomaly Detection Overview}
\label{sec:overview}

\begin{figure*}
    \centering
    \includegraphics[width=0.76\linewidth]{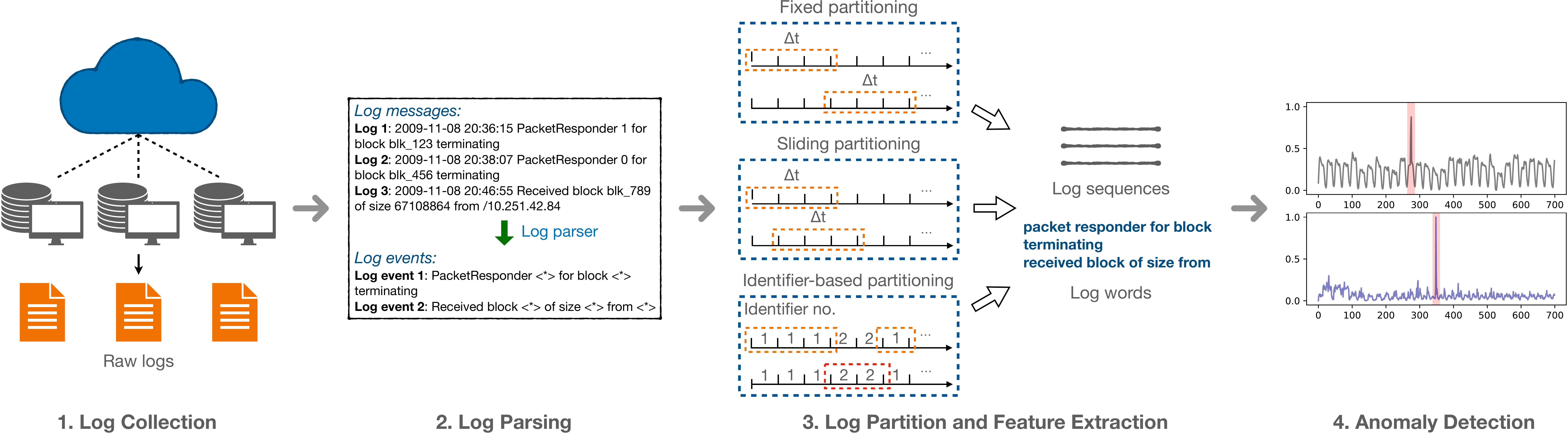}
    \vspace{-0.08in}
    \caption{The Overall Framework of Log-based Anomaly Detection}
    \label{fig:framework}
    \vspace{-0.12in}
\end{figure*}

The overall framework of log-based anomaly detection is illustrated in Fig.~\ref{fig:framework}, which mainly consists of four phases, i.e., \textit{log collection}, \textit{log parsing}, \textit{feature extraction}, and \textit{anomaly detection}.

\subsection{Log Collection}

Software systems routinely print logs to the system console or designated log files to record runtime status. In general, each log is a line of semi-structured text printed by a logging statement in source code, which usually contains a timestamp and a detailed message (e.g., error symptom, target component, IP address). In large-scale systems such as distributed systems, these logs are often collected. The abundance of log data has enabled a variety of log analysis tasks such as anomaly detection and fault localization~\cite{du2017deeplog,yuan2010sherlog}. However, the large volume of the collected logs is overwhelming the existing troubleshooting system. The lack of labeled data also poses difficulty in the analysis of logs.

\subsection{Log Parsing}

After log collection, raw logs are often semi-structured and need to be parsed into a structured format for downstream analysis. This process is called log parsing~\cite{zhu2019tools}. Specifically, log parsing tries to identify the constant/static part and variable/dynamic part of a raw log line. The constant part is commonly referred to as log event, log template, or log key (we use them interchangeably hereafter); the variable part stores the value of the corresponding parameters (e.g., IP address, thread name, job ID, message ID), which could be different depending on specific runs of the system. For example, in Fig.~\ref{fig:framework} (phase two), a log excerpt collected from Hadoop Distributed File System (HDFS) on Amazon EC2 platform~\cite{xu2009detecting} ``\textit{Received block blk\_789 of size 67108864 from /10.251.42.84}'' is parsed into the log event of ``\textit{Received block $<$*$>$ of size $<$*$>$ from $<$*$>$}'', where all parameters are replaced with the token ``$<$*$>$''. Zhu et al.~\cite{zhu2019tools} evaluated 13 methods for automated log parsing and released a toolset.


\subsection{Log Partition and Feature Extraction}
\label{sec:log_partition}

As logs are textual messages, they need to be converted into numerical features such that ML algorithms can understand them. To this end, each log message is first represented with the log template identified by a log parser. Then, log timestamp
and log identifier (e.g., task/job/session ID) are often employed to partition logs into different groups, each representing a log sequence. In particular, timestamp-based log partition usually includes two strategies, i.e., fixed partitioning and sliding partitioning.

\textit{Fixed Partitioning}. Fixed partitioning has a pre-defined partition size, which indicates the time interval used to split the chronologically sorted logs. In this case, there is no overlap between two consecutive fixed partitions. An example is shown in Fig.~\ref{fig:framework} (phase three), where the partition size is denoted as $\Delta t$. $\Delta t$ could be one hour or even one day, depending on the specific problems at hand.


\textit{Sliding Partitioning}. Sliding partitioning has two parameters, i.e., partition size and stride. The stride indicates the forwarding distance of the time window along the time axis to generate log partitions. In general, the stride is smaller than the partition size, resulting in the overlap between different sliding partitions. Therefore, the strategy of sliding partitioning often produces more log sequences than fixed partitioning does, depending on both the partition size and stride. In Fig.~\ref{fig:framework} (phase three), the partition size is $\Delta t$, while the stride is $\Delta t/3$.


\textit{Identifier-based Partitioning}. Identifier-based partitioning sorts logs in chronological order and divides them into different sequences. In each sequence, all logs share a unique and common identifier, indicating they originate from the same task execution. For instance, HDFS logs employ \textit{block id} to record the operations associated with a specific block, e.g., allocation, replication, and deletion. Particularly, log sequences generated in this manner often have varying lengths. For example, sequences with a short length could be due to early termination caused by abnormal execution.


After log partition, many traditional ML-based methods~\cite{he2016experience} generate a vector of log event count as the input feature, where each dimension denotes a log event, and the value counts its occurrence in a log sequence. Different from them, DL-based methods often directly consume the log event sequence. In particular, each element of the sequence can be simply the index of the log event or a more sophisticated feature such as a log embedding vector. The purpose is to learn logs' semantics to make more intelligent decisions. Specifically, the words in a log event are first represented by word embeddings learned with \textit{word2vec} algorithms, e.g., FastText~\cite{joulin2016fasttext} and GloVe~\cite{pennington2014glove}. Then, the word embeddings are aggregated to compose the log event’s semantic vector~\cite{zhang2019robust}.


\subsection{Anomaly Detection}

Based on the log features constructed in the last phase, anomaly detection can be performed, which is to identify anomalous log instances (e.g., logs printed by interruption exceptions). Many traditional ML-based anomaly detectors~\cite{he2016experience} produce a prediction (i.e., an anomaly or not) for the entire log sequence based on its log event count vector. Different from them, many DL-based methods first learn normal log patterns and then determine the normality for each log event. Thus, DL-based methods are capable of locating the exact log event(s) that contaminate the log event sequence, improving the interpretability.

%% file: secs/deep_loglizer.tex
\section{Log Anomaly Detection}
\label{sec:deep_loglizer}

To leverage neural networks for log anomaly detection, the network architecture as well as its loss function should be properly decided. Particularly, the loss function guides how the model learns the log patterns.
In this section, we first elaborate on how existing work formulates the model loss. Then, we introduce six state-of-the-art methods, including four unsupervised methods (i.e., DeepLog~\cite{du2017deeplog}, LogAnomaly~\cite{meng2019loganomaly}, Logsy~\cite{nedelkoski2020self}, and Autoencoder~\cite{farzad2020unsupervised}) and two supervised methods (i.e., LogRobust~\cite{zhang2019robust} and CNN~\cite{lu2018detecting}).

\subsection{Loss Formulation}

The task of log anomaly detection is to uncover anomalous samples in a large volume of log data. A loss should be set for a model with respect to the characteristics of the log data, which serves as the goal to optimize. Generally, each neural network has its typical loss function(s). However, we can set a different goal for it with proper modification in its architecture (e.g.,~\cite{zhang2019robust}). We have summarized the following three representative types of model losses.

\subsubsection{Forecasting Loss}
\label{sec:forecasting_loss}

Forecasting loss guides the model to predict the next appearing log event based on previous observations. A fundamental assumption behind an unsupervised method is that the logs produced by a system's normal executions often exhibit certain stable patterns. When failures happen, such normal log patterns may be violated. For example, erroneous logs appear, the order of log events shifts incorrectly, log sequences become incomplete due to early termination, etc. Therefore, by learning log patterns from normal executions, the method can automatically detect anomalies when the log pattern deviates from normal cases. Specifically, for a log event $e_i$ which shows up at time step $t$, an input window $\mathcal{W}$ is first composed, which contains $m$ log events preceding $e_i$, i.e., $\mathcal{W}=[e_{t-m}, \dots, e_{t-2}, e_{t-1}]$. This is done by dividing log sequences (generated by some log partition strategy) into smaller subsequences. The division process is controlled by two parameters called window size and step size, which are similar to the partition size and stride of the sliding partitioning (Sec.~\ref{sec:log_partition}). A model is then trained to learn a conditional probability distribution $P(e_t=e_i|\mathcal{W})$ for all $e_i$ in the set of distinct log events $E=\{e_1, e_2, \dots, e_n\}$~\cite{du2017deeplog}. In the detection stage, the trained model makes a prediction for a new input window, which will be compared against the actual log event. An anomaly is alerted if the ground truth is not one of the most $k$ probable log events predicted by the model. A smaller $k$ imposes more demanding requirements on the model's performance.

\subsubsection{Reconstruction Loss}

Reconstruction loss is mainly used in autoencoders, which trains a model to copy its input to its output. Specifically, given an input window $\mathcal{W}$ and the model's output $\mathcal{W}^{'}$, the reconstruction loss can be calculated as $sim(\mathcal{W}, \mathcal{W^{'}})$, where $sim$ is a similarity function such as the Euclidean norm. By allowing the model to see normal log sequences, it will learn how to properly reconstruct them. However, when faced with abnormal samples, the reconstruction may not go well, leading to a large loss.

\subsubsection{Supervised Loss}

Supervised loss requires anomaly labels to be available beforehand. It drives the model to automatically learn the features that can help distinguish abnormal samples from normal ones. Specifically, given an input window $\mathcal{W}$ and its label $y_w$, a model is trained to maximize a conditional probability distribution $P(y=y_w|\mathcal{W})$. Commonly-used supervised losses include cross-entropy and mean squared error.

\subsection{Existing Methods}

In this section, we introduce six existing methods, which utilize 
popular neural networks to conduct log-based anomaly detection. They have a particular choice of the model loss and whether to employ the semantic information of logs. We would like to emphasize different combinations (with respect to the model's characteristics and the problem at hand) would yield different methods. For example, by incorporating different loss functions, LSTM models can be either unsupervised~\cite{du2017deeplog,meng2019loganomaly} or supervised~\cite{zhang2019robust}; one method uses the index of log events purely may also accept their semantics; model combinations are also possible as demonstrated by Yen et al~\cite{yen2019causalconvlstm}, i.e., a combination of CNN and LSTM.

\subsubsection{Unsupervised Log-based Anomaly Detection}

The selected four unsupervised methods are introduced as follows:

\textbf{DeepLog}. Du et al.~\cite{du2017deeplog} proposed DeepLog, which is the first work to employ LSTM for log anomaly detection. Particularly, the log patterns are learned from the sequential relations of log events, where each log message is represented by the index of its log event. It is also the first work to detect anomalies in a  forecasting-based fashion, which is widely-used in many follow-up studies.


\textbf{LogAnomaly}. To consider the semantic information of logs, Ma et al.~\cite{meng2019loganomaly} proposed LogAnomaly. Specifically, they proposed \textit{template2Vec} to distributedly represent the words in log templates by considering the synonyms and antonyms therein. For example, the representation vector of the word ``down'' and ``up'' should be distinctly different as they own opposite meanings. To this end, \textit{template2Vec} first searches synonyms and antonyms in log templates and then applies an embedding model, dLCE~\cite{nguyen2016integrating}, to generate word vectors. Finally, the template vector is calculated as the weighted average of the word vectors of the words in the template. Similarly, LogAnomaly adopts forecasting-based anomaly detection with an LSTM model. In this paper, we follow this work to evaluate whether log semantics can bring performance gain to DeepLog.


\textbf{Logsy}. Logsy~\cite{nedelkoski2020self} is the first work utilizing the Transformer~\cite{NIPS2017_3f5ee243} to detect anomalies on log data. It is a classification-based method, which learns log representations in a way to better distinguish between normal data from the system of interest and abnormal samples from auxiliary log datasets. The auxiliary datasets help learn a better representation of the normal data while regularizing against overfitting. Similarly, in this work, we employ the Transformer with multi-head self-attention mechanism. The procedure of anomaly detection follows that of DeepLog~\cite{du2017deeplog}, i.e., forecasting-based. Particularly, we use two types of log event sequences: one only contains the indices of log events as in DeepLog~\cite{du2017deeplog}, while the other is encoded with log semantic information as in LogAnomaly~\cite{meng2019loganomaly}.

\textbf{Autoencoder}. Farzad et al.~\cite{farzad2020unsupervised} were the first to employ autoencoder~\cite{rumelhart1985learning} combined with isolation forest~\cite{liu2008isolation} for log-based anomaly detection. The autoencoder is used for feature extraction, while the isolation forest is used for anomaly detection based on the produced features.
In this paper, we employ an autoencoder to learn representation for normal log event sequences. The trained model is able to encode normal log patterns properly. When dealing with anomalous instances, the reconstruction loss becomes relatively large, which serves as an important signal for anomalies.
We also evaluate whether the model performs better with logs' semantics.

\subsubsection{Supervised Log-based Anomaly Detection}

The selected two supervised methods are introduced as follows:

\textbf{LogRobust}. Zhang et al.~\cite{zhang2019robust} observed that many existing studies of log anomaly detection fail to achieve the promised performance in practice. Particularly, most of them carry a closed-world assumption, which assumes: 1) the log data is stable over time; 2) the training and testing data share an identical set of distinct log events. However, log data often contain previously unseen instances due to the evolution of logging statements and log processing noise. To tackle such a log instability issue, they proposed LogRobust to extract the semantic information of log events by leveraging off-the-shelf word vectors, which is one of the earliest studies to consider logs' semantics as done by Meng et al.~\cite{meng2019loganomaly}.


More often than not, different log events have distinct impacts on the prediction result. Thus, LogRobust incorporates the attention mechanism~\cite{bahdanau2014neural} into a Bi-LSTM model to assign different weights to log events, called attentional BiLSTM. Specifically, LogRobust adds a fully-connected layer as the attention layer to the concatenated hidden state $h_t$. It calculates an attention weight (denoted as $a_t$), indicating the importance of the log event at time step $t$ as $a_t=tanh(W_t^a\cdot h_t)$, where \noindent $W_t^a$ is the weight of the attention layer. Finally, LogRobust sums all hidden states at different time steps with respect to the attention weights and employs a softmax layer to generate the classification result (i.e., anomaly or not) as $prediction=softmax(W\cdot (\sum_{t=1}^{T} a_t\cdot h_t))$, where $W$ is the weight of the softmax layer, and $T$ is the length of the log sequence.

\textbf{CNN}. Lu et al.~\cite{lu2018detecting} conducted the first work to explore the feasibility of CNN~\cite{lecun1998gradient} for log-based anomaly detection. The authors first constructed log event sequences by applying identifier-based partitioning (Sec.~\ref{sec:log_partition}), where padding or truncation is applied to obtain consistent sequence lengths. Then, to perform convolution calculation which requires a two-dimensional feature input, the authors proposed an embedding method called \textit{logkey2vec}. Specifically, they first created a trainable matrix whose shape equals $\#distinct~log~events\times embedding~size$ (a tuneable hyperparameter). Then, different convolutional layers (with different shape settings) are applied, and their outputs are concatenated and fed to a fully-connected layer to produce the prediction result.

\subsection{Tool Implementation}

In the literature, tremendous efforts have been devoted to the development of DL-based log anomaly detection. While they have achieved remarkable performance, they have not yet been fully integrated into industrial practices. This gap largely comes from the lack of publicly available tools that are ready for industrial usage. For site reliability engineers who have limited expertise and experience in ML techniques, re-implementation requires non-trivial efforts. Moreover, they are often busy with emerging issue mitigation and resolution. Yet, the implementation of DL models is usually time-consuming, which involves the process of parameter tuning. This motivates us to develop a unified toolkit that provides out-of-the-box DL-based log anomaly detectors.

We implemented the studied six anomaly detection methods in Python with around 3,000 lines of code and packaged them as a toolkit with standard and unified input/output interfaces. Moreover, our toolkit aims to provide users with the flexibility for model configuration, e.g., different loss functions and whether to use logs' semantic information. For DL model implementation, we utilize a popular machine learning library, namely PyTorch~\cite{pytorch}. PyTorch provides basic building blocks (e.g., recurrent layers, convolution layers, Transformer layers) for the construction of a variety of neural networks such as LSTM, CNN, the Transformer, etc. For each model, we experiment with different architecture and parameter settings. We employ the setting that constantly yields a good performance across different log datasets.

%% file: secs/eval.tex
\section{Evaluation}
\label{sec:evaluation}

In this section, we evaluate six DL-based log anomaly detectors on two widely-used datasets~\cite{he2020loghub} and report the benchmarking results in terms of accuracy, robustness, and efficiency. They represent the key quality of interest to consider during industrial deployment.

\textit{Accuracy} measures the ability of a method to distinguish anomalous logs from normal ones. This is the main focus in this field. A large false-positive rate would miss critical system failures, while a large false-negative rate would incur a waste of engineering effort.
    
\textit{Robustness} measures the ability of a method to detect anomalies with the presence of unknown log events. As modern software systems involve rapidly, this issue starts to gain more attention from both academia and industry. One common solution is leveraging logs' semantic information by assembling word-level features.
    
\textit{Efficiency} gauges the speed of a model to conduct anomaly detection. We evaluate the efficiency by recording the time an anomaly detector takes in its training and testing phases. Nowadays, terabytes and even petabytes of data are being generated on a daily basis, imposing stringent requirements on the model's efficiency.

\subsection{Experiment Design}

\subsubsection{Log Dataset}

He et al.~\cite{he2020loghub} released Loghub, a large collection of system log datasets. Due to space limitations, in this paper, we only report results evaluated on two popular datasets, namely, HDFS~\cite{xu2009detecting} and BGL~\cite{oliner2007supercomputers}. Nevertheless, our toolkit can be easily extended to other datasets. Table~\ref{tab:data_statistics} summarizes the dataset statistics.

\textit{HDFS}. HDFS dataset contains 11,175,629 log messages, which are generated by running map-reduce tasks on more than 200 Amazon’s EC2 nodes~\cite{du2017deeplog}. Particularly, each log message contains a unique \textit{block\_id} for each block operation such as allocation, writing, replication, deletion. Thus, identifier-based partitioning can be naturally applied to generate log event sequences. After preprocessing, we end up with 575,061 log sequences, among which 16,838 samples are anomalous. A log sequence will be predicted as anomalous if any of its log windows, $\mathcal{W}$, is identified as an anomaly.
    
\textit{BGL}. BGL dataset contains 4,747,963 log messages, which are collected from a BlueGene/L supercomputer at Lawrence Livermore National Labs. Unlike HDFS, logs in this dataset have no identifier to distinguish different job executions. Thus, timestamp-based partitioning is applied to slice logs into log sequences. The number of the resulting sequences depends on the partition size (and stride).
In the BGL dataset, 348,460 log messages are labeled as failures. A log sequence is marked as an anomaly if it contains any failure logs.




\subsubsection{Evaluation Metrics}

Since log anomaly detection is a binary classification problem, we employ \textit{precision}, \textit{recall}, and \textit{F1 score} for accuracy evaluation. Specifically, precision measures the percentage of anomalous log sequences that are successfully identified as anomalies over all the log sequences that are predicted as anomalies: $precision = \frac{TP}{TP + FP}$; recall calculates the portion of anomalies that are successfully identified by a model over all the actual anomalies: $recall = \frac{TP}{TP + FN}$; F1 score is the harmonic mean of precision and recall: $F1~score = 2 \times \frac{precision \times recall}{precision + recall}$. \textit{TP} is the number of anomalies that are correctly disclosed by the model, \textit{FP} is the number of normal log sequences that are wrongly predicted as anomalies by the model, \textit{FN} is the number of anomalies that the model misses.

\subsubsection{Experiment Setup}

For a fair comparison, all experiments are conducted on a machine with 4 NVIDIA Titan V Pascal GPUs (12GB of RAM), 20 Intel(R) Xeon(R) Gold 6148 CPU @ 2.40GHz, and 256GB of RAM. The parameters of all methods are fine-tuned to achieve the best results. To avoid bias from randomness, we run each method five times and report the best result.

For all datasets, we first sort logs in chronological order and apply log partition to generate log sequences, which will then be shuffled. Note we do not shuffle the input windows, $\mathcal{W}$, generated from log sequences. Next, we utilize the first 80\% of data for model training and the remaining 20\% for testing. Particularly, for unsupervised methods that require no anomalies for training, we remove them from the training data. This is because many unsupervised methods try to learn the normal log patterns and alert anomalies when such patterns are violated. Thus, they require anomaly-free log data to yield the best performance. Nevertheless, we will evaluate the impact of anomaly samples in training data. For log partition, we apply identifier-based partitioning to HDFS and fixed partitioning with six hours of partition size to BGL. The default values of window size and step size are ten and one, which are set empirically based on our experiments. For HDFS and BGL, we set $k$ as ten and 50, respectively.
In particular, a log event sequence will be regarded as an anomaly if any one of its log windows, $\mathcal{W}$, is predicted as anomalous.

\subsection{Accuracy of Log Anomaly Detection}

In this section, we explore models' accuracy. We first show the results when log event sequences are composed of log events' indices. Then, we evaluate the effectiveness of logs' semantics by incorporating it into the log sequences. Finally, we control the ratio of anomalies in the training data to see its influence.

\begin{table}[t]
\begin{center}
\caption{Dataset Statistics}
\vspace{-0.1in}
\label{tab:data_statistics}
\normalsize
 \begin{tabular}{c|c|c|c}
  \toprule
  
  \textbf{Dataset} & \textbf{Time span} & \textbf{\#Logs} & \textbf{\#Anomalies}\\
  
  \toprule[0.8pt]
  
HDFS & 38.7 hrs & 11,175,629 & 16,838 \\
BGL & 7 mos & 4,747,963 & 348,460 \\

  \bottomrule
 \end{tabular}
\vspace{-0.1in}
\end{center}
\end{table}

\begin{table*}[t]
\begin{center}
\caption{Accuracy of DL-based Log Anomaly Detection Methods}
\vspace{-0.1in}
\label{tab:acc_semantics}
\normalsize
 \begin{tabular}{c|c|c|c|c|c|c}
  \toprule
  & \multicolumn{3}{c|}{HDFS (w/o and w/ semantics)} & \multicolumn{3}{c}{BGL (w/o and w/ semantics)} \\
  
  \textbf{Models} & \textbf{Precision} & \textbf{Recall} & \textbf{F1 score} & \textbf{Precision} & \textbf{Recall} & \textbf{F1 score} \\
  
  \toprule[0.8pt]
  
  LSTM~\cite{du2017deeplog,meng2019loganomaly} & \textbf{0.96}/\textbf{0.965} & \textbf{0.928}/0.904 & \textbf{0.944}/\textbf{0.945} & \textbf{0.935}/\textbf{0.946} & \textbf{0.989}/0.989 & \textbf{0.961}/\textbf{0.967} \\
  Transformer~\cite{nedelkoski2020self} & 0.946/0.86 & 0.867/\textbf{1.0} & 0.905/0.925 & \textbf{0.935}/0.917 & 0.977/\textbf{1.0} & 0.956/0.957 \\
  Autoencoder~\cite{farzad2020unsupervised} & 0.881/0.892 & 0.878/0.869 & 0.88/0.881 & 0.791/0.942 & 0.773/0.92 & 0.782/0.931 \\

  \midrule

  Attn. BiLSTM~\cite{zhang2019robust} & 0.933/0.934 & 0.989/\textbf{0.995} & 0.96/0.964 & \textbf{0.989}/0.989 & \textbf{0.977}/\textbf{0.977} & \textbf{0.983}/0.983 \\
  CNN~\cite{lu2018detecting} & \textbf{0.946}/\textbf{0.943} & \textbf{0.995}/\textbf{0.995} & \textbf{0.97}/\textbf{0.969} & 0.966/\textbf{1.0} & \textbf{0.977}/\textbf{0.977} & 0.972/\textbf{0.989} \\

  \bottomrule
 \end{tabular}
\vspace{-0.08in}
\end{center}
\end{table*}

\subsubsection{Accuracy without Log Semantics}

The performance of different methods is shown in Table~\ref{tab:acc_semantics} (the first figures). It is not surprising that supervised methods generally achieve better performance than their unsupervised counterparts do. For HDFS and BGL, the best F1 scores (hereafter, we mainly talk about this metric unless otherwise stated) that unsupervised methods can attain are 0.944 and 0.961, respectively, both of which come from the LSTM model~\cite{du2017deeplog}. On the other hand, supervised methods have pushed them to 0.97 (by CNN~\cite{lu2018detecting}) and 0.983 (by attentional BiLSTM~\cite{zhang2019robust}), achieving noticeable improvements. Among all unsupervised methods, Autoencoder, which is the only construction-based model, performs relatively poorly, i.e., 0.88 in HDFS and 0.782 in BGL. Nevertheless, it possesses the merit of great resistance against anomalies in training data, as we will show later. LSTM shows outstanding overall performance, demonstrating its exceptional ability in capturing normal log patterns. On the supervised side, CNN and attentional BiLSTM achieve comparable results in both datasets, outperforming unsupervised methods by around 2\%.

We also present the results of traditional ML-based methods in Table~\ref{tab:traditional_ml} using the toolkit released by He et al.~\cite{he2016experience}, which contains three unsupervised methods, i.e., Log Clustering (LC), Principal Component Analysis (PCA), Invariant Mining (IM), and three supervised methods, i.e., Logistic Regression (LR), Decision Tree (DT), and Support Vector Machine (SVM). For HDFS, Decision Tree achieves a remarkable performance, i.e. 0.998, ranking the best among all. Other traditional ML-based methods are generally defeated by their DL-based counterparts. This is also the case for BGL. Moreover, traditional unsupervised methods seem to be inapplicable for BGL, e.g., the F1 score of PCA is only 0.56, while unsupervised DL-based methods yield much better results. Particularly, compared with the experiments conducted by He et al.~\cite{he2016experience}, we achieve better results on BGL when running both DL-based and traditional ML-based methods. This attributes to the fact that we apply shuffling to the dataset, which alleviates the issue of unseen logs in BGL's testing data. Note that this is done at the level of log sequences. The order of log events in each input window is preserved.

\begin{table}[t]
\begin{center}
\caption{Accuracy of Traditional ML-based Methods}
\vspace{-0.1in}
\label{tab:traditional_ml}
\normalsize
 \begin{tabular}{c|c|c|c|c|c|c}
  \toprule
  & \multicolumn{3}{c|}{HDFS} & \multicolumn{3}{c}{BGL} \\
  
  \textbf{Meth.} & \textbf{Prec.} & \textbf{Rec.} & \textbf{F1} & \textbf{Prec.} & \textbf{Rec.} & \textbf{F1} \\
  
  \toprule[0.9pt]
  
  LC & \textbf{1.0} & 0.728 & 0.843 & \textbf{0.975} & 0.443 & 0.609 \\
  PCA & 0.971 & 0.628 & 0.763 & 0.52 & \textbf{0.619} & 0.56 \\
  IM & 0.895 & \textbf{1.0} & \textbf{0.944} & 0.86 & 0.489 & \textbf{0.623} \\

  \midrule

  LR & 0.95 & 0.921 & 0.935 & 0.791 & 0.818 & 0.804 \\
  DT & \textbf{0.997} & \textbf{0.998} & \textbf{0.998} & 0.964 & \textbf{0.92} & 0.942 \\
  SVM & 0.956 & 0.913 & 0.934 & \textbf{0.988} & 0.909 & \textbf{0.947} \\

  \bottomrule
 \end{tabular}
\vspace{-0.1in}
\end{center}
\end{table}

\subsubsection{Accuracy with Log Semantics}

To leverage logs' semantics, some work, e.g.,~\cite{zhang2019robust}, adopts off-the-shelf word vectors, e.g., pre-trained on Common Crawl Corpus dataset using the FastText algorithm~\cite{joulin2016fasttext}. Different from them, in our experiments, we randomly initialize the embedding vector for each word as we did not observe much improvement when following their configurations. An important reason is that many words in logs are not covered in the pre-trained vocabulary. Table~\ref{tab:acc_semantics} (the second figures) presents the performance when models have access to logs' semantic information for anomaly detection. We can see almost all methods benefit from logs' semantics, e.g., Autoencoder obtains nearly 15\% of performance gain. Particularly, the best F1 scores achieved by unsupervised and supervised methods on BGL dataset become 0.967 (by LSTM~\cite{du2017deeplog}) and 0.989 (by CNN~\cite{lu2018detecting}), respectively, while the best F1 scores on HDFS dataset remain almost unchanged. Nevertheless, Decision Tree is still undefeated on HDFS dataset. Logs' semantics not only promotes the accuracy of anomaly detection but also brings other kinds of benefits to the models, as we will show in the next sections.

\begin{flushleft}
\begin{tabular}{|>{\raggedright}p{0.95\columnwidth}|}
\noalign{\vskip4pt}
\hline 
\textsc{Finding 1}. Supervised methods generally achieve superior per- formance than unsupervised methods do. Logs' semantics in- deed contributes to the detection of anomalies, especially for unsupervised methods. \tabularnewline
\hline 
\multicolumn{1}{>{\raggedright}p{0.95\columnwidth}}{}\tabularnewline
\noalign{\vskip-6pt}
\end{tabular}
\par\end{flushleft}

\subsubsection{Accuracy with Varying Anomaly Ratio}

\begin{figure}[t]
    \centering
    \includegraphics[width=0.8\linewidth]{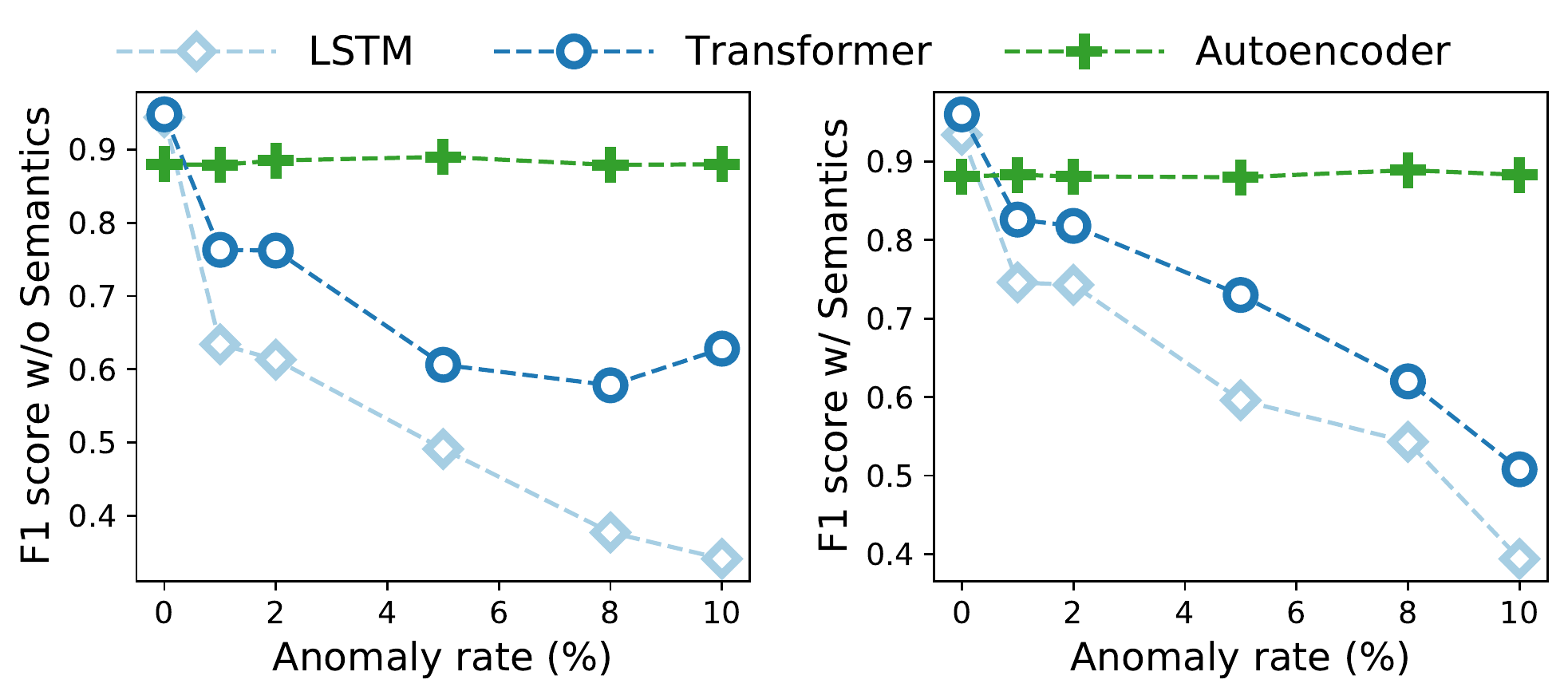}
    \vspace{-0.1in}
    \caption{Accuracy w/ varying anomaly ratio in training data}
    \label{fig:acc_with_train_anomaly_rate}
    \vspace{-0.1in}
\end{figure}

In this experiment, we evaluate how the anomalies in training data will impact the performance of unsupervised DL-based methods. The motivation is that some works claim that a small amount of noise (i.e., anomalous instances) in training data only has a trivial impact on the results. This is because normal data are dominant, and the model will forget the anomalous patterns. In our previous experiments, we remove all anomalies from the training data such that the normal patterns can be best learned. However, in reality, anomalies are inevitable. We simulate this situation by randomly putting a specific portion of anomalies (from 1\% to 10\%) back to the training data. The results on HDFS are shown in Fig.~\ref{fig:acc_with_train_anomaly_rate}, where we experiment without and with logs' semantics. Clearly, even with just 1\% of anomalies, the F1 score of both LSTM and the Transformer drops significantly to 0.634 and 0.763, respectively. Logs' semantics safeguards around 10\% of performance against the anomalies. When the percentage of anomalies reaches 10\%, the F1 score of LSTM even degrades to less than 0.4. Interestingly, Autoencoder exhibits great resilience against noisy training data, which demonstrates that compared with forecasting-based methods, construction-based methods are indeed able to forget the anomalous log patterns.

\begin{flushleft}
\begin{tabular}{|>{\raggedright}p{0.95\columnwidth}|}
\noalign{\vskip4pt}
\hline 
\textsc{Finding 2}. For forecasting-based methods, anomalies in train- ing data can quickly deteriorate the performance. Different from them, reconstruction-based methods are more resistant to training data containing anomalies. \tabularnewline
\hline 
\multicolumn{1}{>{\raggedright}p{0.95\columnwidth}}{}\tabularnewline
\noalign{\vskip-6pt}
\end{tabular}
\par\end{flushleft}

\subsection{Robustness of Log Anomaly Detection}

In this section, we study the robustness of the selected anomaly detectors, i.e., the accuracy with the presence of unseen logs. We also compare them against traditional ML-based methods. To simulate the log instability issue, we follow Zhang et al.~\cite{zhang2019robust} to synthesize new log data. Given a randomly sampled log event sequence in the testing data, we apply one of the following four noise injection strategies: randomly injecting a few pseudo log events (generated by trivial word addition/removal or synonym replacement) or deleting/shuffling/duplicating a few existing log events in the sequence. We inject the synthetic log sequences into the original log data according to a specific ratio (from 5\% to 20\%). With the injected noises, DL-based methods that leverage logs' semantics can continue performing anomaly prediction without retraining. However, their traditional ML-based counterparts need to be retrained because the number of distinct log events is fixed. We follow Zhang et al.~\cite{zhang2019robust} to append an extra dimension to the log count vector (for both training and testing data) to host all pseudo log events.

\begin{figure}[t]
    \centering
    \includegraphics[width=0.81\linewidth]{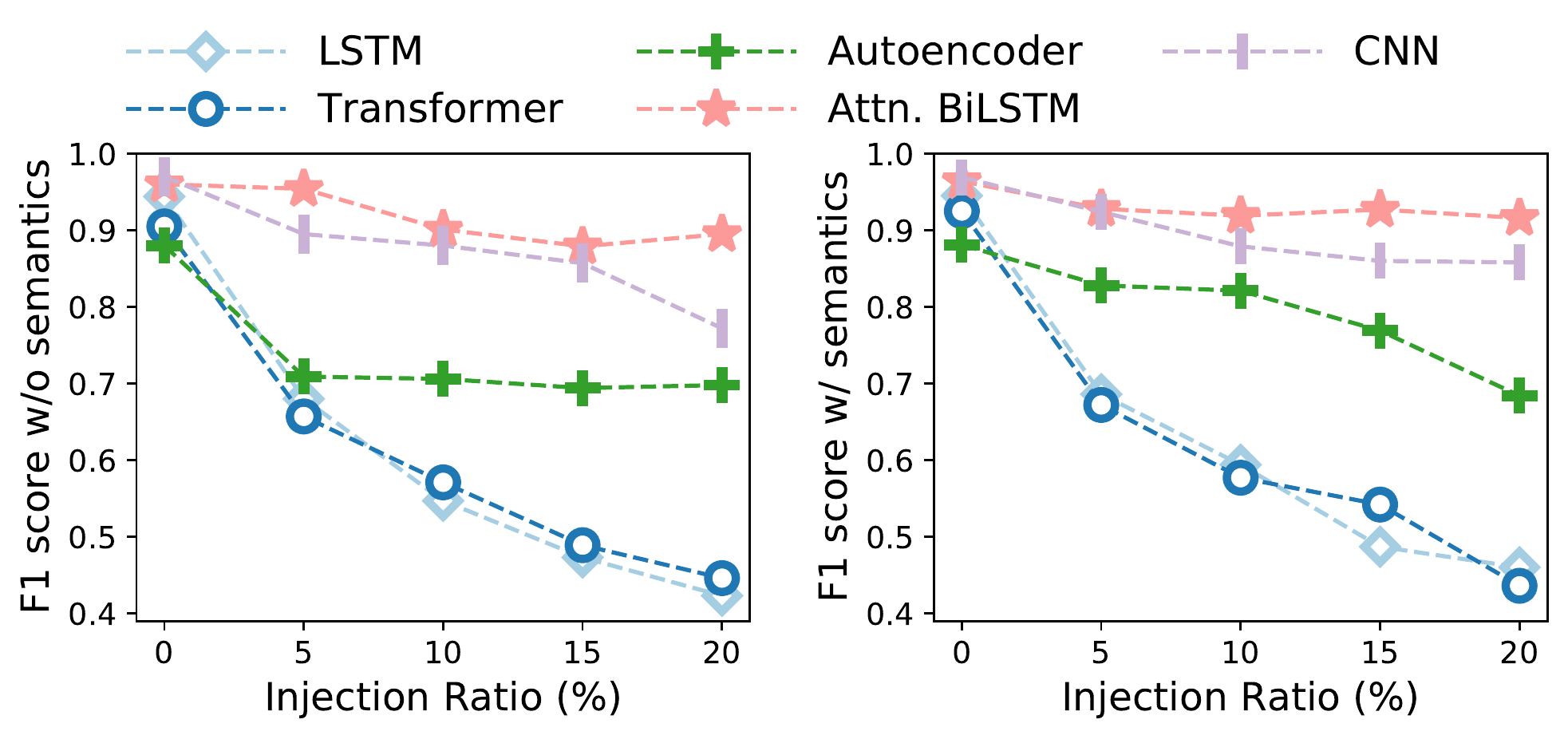}
    \vspace{-0.1in}
    \caption{Robustness of DL-based methods on HDFS}
    \label{fig:robustness_with_inject_ratio}
    \vspace{-0.1in}
\end{figure}

\begin{figure}[t]
    \centering
    \includegraphics[width=0.81\linewidth]{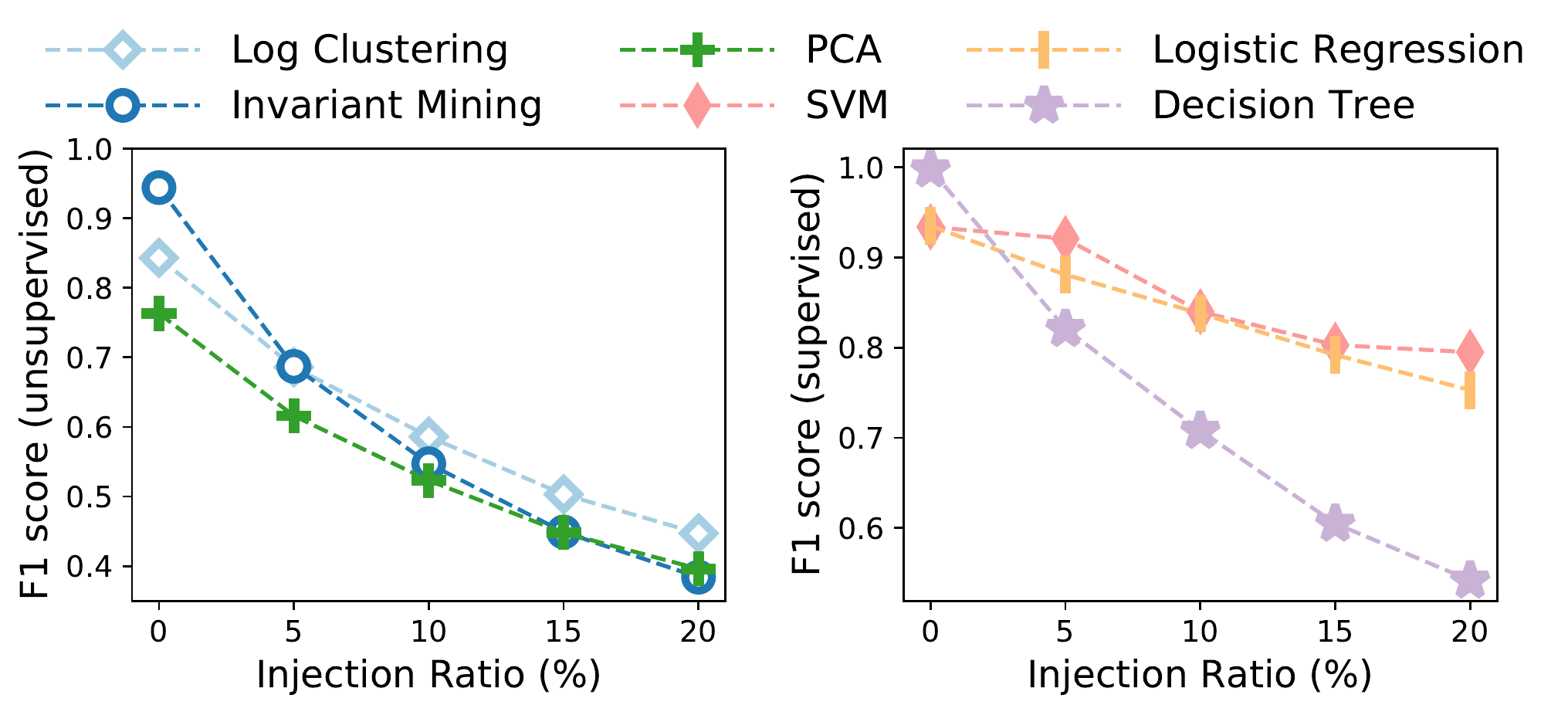}
    \vspace{-0.1in}
    \caption{Robustness of ML-based methods on HDFS}
    \label{fig:robustness_loglizer}
    \vspace{-0.1in}
\end{figure}

The results of DL-based methods on HDFS are presented in Fig.~\ref{fig:robustness_with_inject_ratio}. Clearly, the performance of all models is harmed by the injected noises. In particular, unsupervised methods are much more vulnerable than supervised methods. For LSTM and the Transformer, 5\% of noisy logs suffice to degrade their F1 score by more than 20\%. Logs' semantics offers little help in this case. Autoencoder again demonstrates good robustness against noise and benefits more from logs' semantics. The situations of supervised models are much better. With the access to logs' semantics, they successfully maintain an F1 score of around 0.9 even with 20\% noises injected, while that of LSTM and the Transformer are both lower than 0.5. This proves that logs' semantic information indeed helps DL-based models adapt to unprecedented log events. On the side of traditional ML-based methods in Fig.~\ref{fig:robustness_loglizer}, unsupervised methods are also more sensitive than their supervised counterparts. In particular, SVM and Logistic Regression achieve the best performance, i.e., around 0.8 of the F1 score retained when the testing data contains 20\% noises. Under the same setting, PCA and Invariant Mining have the worst results, i.e., around 0.4 of F1 score.

\begin{flushleft}
\begin{tabular}{|>{\raggedright}p{0.95\columnwidth}|}
\noalign{\vskip4pt}
\hline 
\textsc{Finding 3}. Unprecedented logs have a significant impact on anomaly detection. Supervised methods exhibit better robust- ness against such logs than unsupervised methods. Moreover, logs' semantics can further promote the robustness. \tabularnewline
\hline 
\multicolumn{1}{>{\raggedright}p{0.95\columnwidth}}{}\tabularnewline
\noalign{\vskip-6pt}
\end{tabular}
\par\end{flushleft}

\subsection{Efficiency of Log Anomaly Detection}

\begin{figure}[t]
    \centering
    \includegraphics[width=0.8\linewidth]{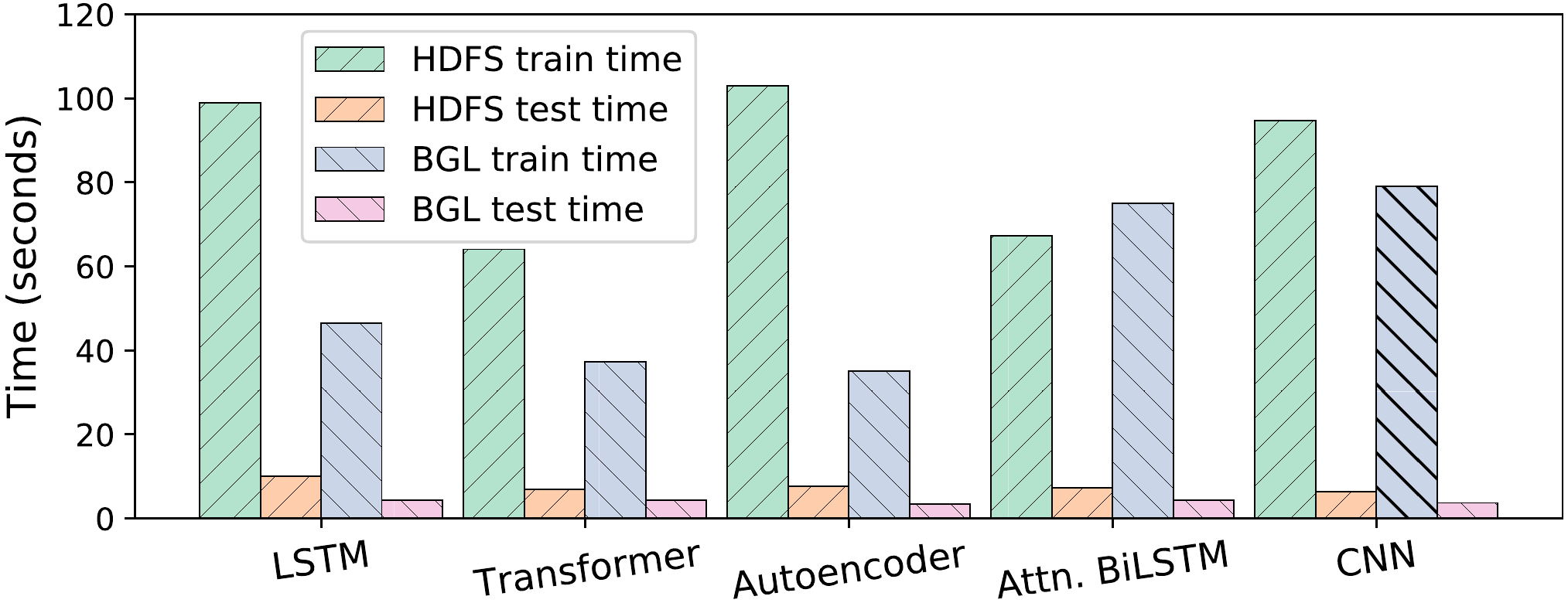}
    \vspace{-0.1in}
    \caption{Efficiency on both HDFS and BGL datasets}
    \label{fig:efficiency_wo_semantics}
    \vspace{-0.15in}
\end{figure}

In this section, we evaluate the efficiency of different models by recording the time spent on the training and testing phases for both datasets. The results are given in Fig.~\ref{fig:efficiency_wo_semantics}, where we do not consider logs' semantics. We can see each model generally requires tens of seconds for model training and around five seconds for testing. BGL consumes less time due to its smaller volume. For HDFS, LSTM and Autoencoder are the most time-consuming models for training; while for BGL, supervised models demand more time. On the other hand, some traditional ML-based methods, i.e., Logistic Regression, Decision Tree, SVM, and PCA show superior performance over DL-based models, which only take seconds for model training. SVM and PCA can even produce results in a real-time manner. However, Invariant Mining consumes thousands of seconds for pattern mining on HDFS. Regarding model testing, except for Log Clustering, other methods only require tens of milliseconds.


\begin{flushleft}
\begin{tabular}{|>{\raggedright}p{0.95\columnwidth}|}
\noalign{\vskip4pt}
\hline 
\textsc{Finding 4}. Compared to traditional ML-based methods, DL- based methods often require more time for training and testing. Some ML-based methods demonstrate outstanding efficiency. \tabularnewline
\hline 
\multicolumn{1}{>{\raggedright}p{0.95\columnwidth}}{}\tabularnewline
\noalign{\vskip-6pt}
\end{tabular}
\par\end{flushleft}

%% file: secs/industry.tex
\section{Industrial Practices}
\label{sec:industry}


In this section, we present an industrial case study of deploying automated log-based anomaly detection in production at Huawei Cloud. The model is an optimized version based on 
DeepLog~\cite{du2017deeplog}.
DeepLog was selected for its simplicity and superior performance (see Table~\ref{tab:acc_semantics}). It is one of the most highly-cited papers in the field and serves as the prototype for many follow-up models. Services in Huawei Cloud serve hundreds of millions of users worldwide and produce terabytes of log data on a daily basis. Such a large volume of data is impractical for engineers to detect anomalies manually. Thus, automated log anomaly detection is in high demand.

To share our industrial practices, we first introduce the deployment architecture of our model. Then, we summarize the real-world challenges that we met during its usage in production. Finally, we discuss some promising future improvements that can push this field forward beyond current practices.

\subsection{Industrial Deployment}

\begin{figure}[t]
    \centering
    \includegraphics[width=0.9\linewidth]{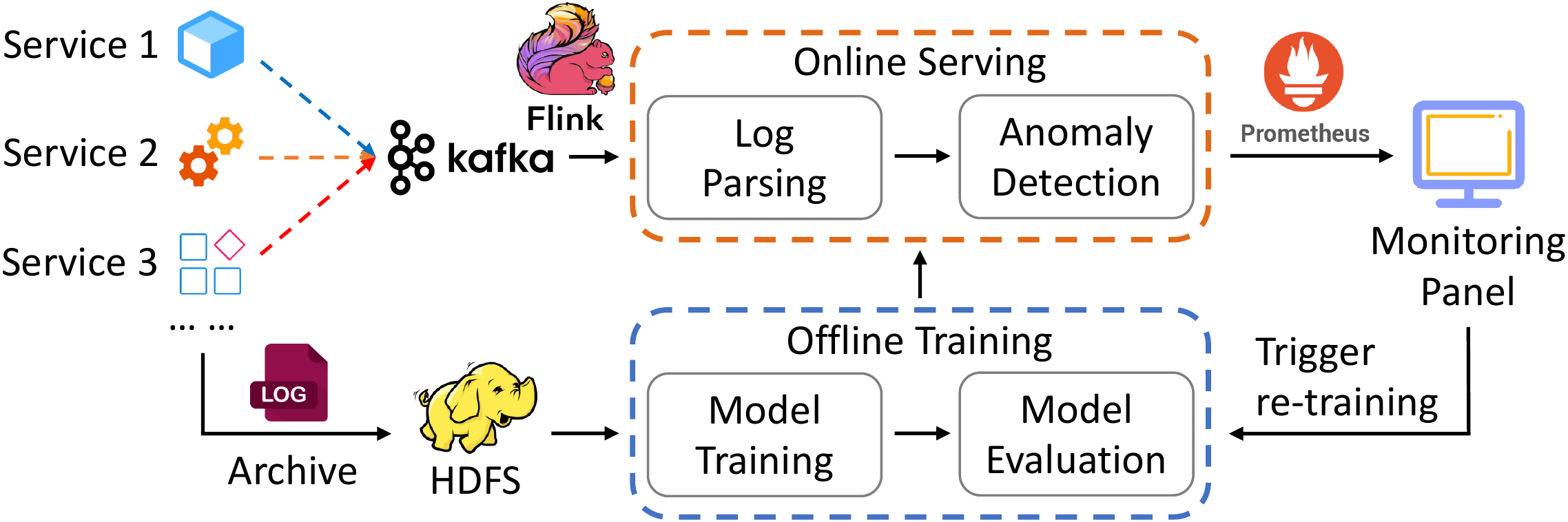}
    \vspace{-0.08in}
    \caption{Log anomaly detection pipeline in production}
    \label{fig:deployment}
    \vspace{-0.2in}
\end{figure}

Fig.~\ref{fig:deployment} illustrates the deployment architecture of our log anomaly detection pipeline in production, which consists of two stages, i.e., offline training and online serving. In the online stage. We employ Kafka as a streaming channel for online log analytics. Data producers are different services that generate raw log data at runtime. Each service corresponds to one Kafka topic for data streaming. Our model acts as the data consumer and performs anomaly detection for each service. We adopt Apache Flink for distributed log preprocessing and anomaly detection, which can process streaming data with high performance and low latency. The detection results are visualized on a monitoring panel through Prometheus. Engineers will confirm true anomalies or flag false positives with simple clicks. In the offline stage, the model is (re)trained. Specifically, raw logs are first archived and maintained in Apache HDFS. Then, they can be retrieved for model (re)training and evaluation. A threshold needs to be set manually for alerting anomalies. In particular, engineers can trigger model retraining if they see performance degradation on the monitoring panel.

\subsection{Real-World Challenges}

Our model has been deployed in Huawei Cloud for 14 months. During this period, we have seen it shedding lights on automated log-based anomaly detection and bringing benefits to online services by reporting high-risk anomalies. We select a large-scale service system and retrieve over one year of anomalies predicted for it. The result indicates that more than 85\% of them have been confirmed by engineers. However, we have also noticed its inadequacy in combating the ever-increasing complexity of cloud systems. Based on our experience of model deployment and operations, we have summarized the following challenges in the way of pursuing highly intelligent and reliable log-based anomaly detection.


\textit{High Complexity}. Compared to HDFS and BGL, production logs present much greater complexity. For example, logs can be interleaving~\cite{nandi2016anomaly,he2021survey} due to concurrent execution of multiple tasks, the number of log events is overwhelming the model, etc.

\textit{Threshold Re-determination}. We found that the threshold for anomaly assertion requires adjustment from time to time after the model has been deployed. The optimal setting obtained in offline training may not always be applicable for online deployment.

\textit{Concept Drift}. Modern software systems continuously undergo feature upgrades, which may incur concept drift. For example, new log events may emerge, log patterns may change. Without proper online learning capability, frequent model retraining is required.

\textit{Large-Volume and Low-Quality Data}. Although a massive amount of log data has been produced, a significant portion of logs only records plain system runtime information. Moreover, there is currently a lack of rigorous guides and specifications on developer logging behaviors. Thus, logs exhibit different styles across different service teams and often contain meaningless tokens, making it hard to design an appropriate log parser for feature extraction.

\textit{Unsatisfactory Interpretability}. Although DL-based methods possess better interpretability, e.g., specifically locate the problematic logs, they are still unable to explain the occurrence of anomalies from system perspective.
Learning logs' semantics can be a promising direction to further enhance interpretability. However, many logs themselves are meaningless. Tokens like abbreviations and self-defined words are hardly understandable to other engineers.

\textit{Incorrect Model Strategy}. An important design goal of DeepLog is to learn log patterns that reflect programs' rigorous set of logic and control flows~\cite{du2017deeplog}. However, most anomalies stem from the occurrence of particular error logs instead of incorrect orders of log events. In this case, maintaining a set of important log events
may be sufficient for anomaly detection. To make more accurate anomaly detection, multiple strategies may need to be integrated.

\textit{Others}. We also recognize labeling and log privacy issues. Labeling issue refers to the deficiency of anomaly labels for model evaluation. Engineers reported that many cases are too vague for normality determination. Due to privacy policies, it is prohibitive to access some customers' logs. Without some detailed log information, prompt anomaly detection becomes a greater challenge.

\subsection{Future Improvements}


\subsubsection{Closer Engineering Collaboration}

Companies need to establish and align on a clear objective at executive level. The infrastructure development, service architecture design, engineers' mindsets should serve this clear objective collaboratively. In current practices, logs are collected from different services in an ad-hoc manner and used for independent model development. A pipeline of log data generation, collection, labeling, and usage should be built. In this process, the engineering principles should include data/label sanity check and continuous model-quality validation.

\subsubsection{Better Logging Practices}

Good-quality log data play an essential role in many downstream log analytics tasks, including anomaly detection, failure diagnosis, performance optimization, user profiling, etc. Thus, better logging practices should be established to guide the writing of logging statements. Some examples are: 1) a logging statement should include a timestamp, a proper verbosity level, etc.; 2) the context should be made clear by including critical variables, components, process IDs, etc.; 3) a log message should be meaningful and understandable to its audience; 5) apply template-based logging~\cite{message_templates} that is human-friendly and machine-readable; 6) the number of logging statements should be kept in a proper level.

\subsubsection{Model Improvement}

The challenges of log-based anomaly detection for today's software systems also pose requirements on the model's capabilities. Promising evolution directions include online learning, human knowledge incorporation (e.g., human-in-the-loop design), multi-source learning (i.e., combine other system monitoring data such as metrics and incident tickets~\cite{chen2020towards}), etc. A recent study~\cite{zhao2021empirical} reveals several limitations that end-to-end solutions suffer from in practice. The authors argue that multiple aspects of logs (e.g., keywords, log event count/sequence) should be addressed. They proposed a generic log anomaly detection system based on ensemble learning. Exploring the semantic relations between log events is also an appealing direction. Specifically, some logs may describe different steps of an operation, e.g., file sending and file receiving. Being able to discover and analyze such relations 
is crucial for accurate anomaly detection and automated fault localization.



%% file: secs/related_work.tex
\section{Related Work}
\label{sec:related_work}

\textbf{Log analysis}. Logs have become imperative in the assurance of software systems' reliability and continuity because they are often the only data available that record software runtime information. Typical applications of logs include anomaly detection~\cite{he2016experience,xu2009detecting,du2017deeplog}, failure prediction~\cite{russo2015mining,sahoo2003critical}, failure diagnosis~\cite{yuan2010sherlog,zhou2019latent}, etc. Most log analysis studies involve two main steps, i.e., log parsing and log mining. Based on whether log parsing can be conducted in a streaming manner, log parsers can be categorized into offline and online. Zhu et al.~\cite{zhu2019tools} conducted a comprehensive evaluation study on 13 automated log parsers and reported the benchmarking results.
More recently, Dai et al.~\cite{dai2020logram} proposed an online parser called Logram, which considers the n-grams of logs. The core insight is 
that frequent n-grams are more likely to be part of log templates.

Many efforts have also been devoted to log mining, especially anomaly detection due to its practical significance. They can be roughly categorized into two classes as studied in this paper, i.e., traditional ML-based methods and DL-based methods. For example, Xu et al.~\cite{xu2009detecting} were the first to apply PCA to mine system problems from console logs. By mining invariants among log messages, Lou et al.~\cite{lou2010mining} detected system anomalies when any of the invariants are violated. Lin et al.~\cite{lin2016log} proposed LogCluster, which recommends representative log sequences for problem identification by clustering similar log sequences. He et al.~\cite{he2018identifying} proposed Log3C to incorporate system KPIs into the identification of high-impact issues in service systems. Some work~\cite{fu2009execution,nandi2016anomaly} employs graph models such as finite state machines and control flow graphs to capture a system's normal execution paths. Anomalies are alerted if the transition probability or sequence violates the learned graph model.

In recent years, there has been a growing interest in applying neural networks to log anomaly detection. For example, Du et al.~\cite{du2017deeplog} proposed DeepLog, which is the first work to adopt an LSTM to detect log anomalies in an unsupervised manner. Meng et al.~\cite{meng2019loganomaly} proposed LogAnomaly to extend their work by incorporating logs' semantics. To address the issue of log instability, i.e., new logs may emerge during system evolution, Zhang et al.~\cite{zhang2019robust} proposed a supervised method called LogRobust, which also considers logs' semantic information. Wang et al.~\cite{yang2021semi} addressed the issue of insufficient labels via probabilistic label estimation and designed an attention-based GRU neural network. Lu et al.~\cite{lu2018detecting} explored the feasibility of CNN for this task. Other models include the Transformer~\cite{nedelkoski2020self}, LSTM-based generative adversarial network~\cite{xia2021loggan}, etc.

\textbf{Empirical studies on logs}. Empirical studies are also an important topic in the log analysis community, which derives valuable insights from abundant research work in the literature and industrial practices. For example, Yuan et al.~\cite{yuan2012conservative} studied the logging practices of open-source systems and provided developers with suggestions for improvement. Fu et al.~\cite{fu2014developers,zhu2015learning} focused on the logging practices on the industry side. The work done by He et al.~\cite{he2016experience} is the most related study to ours, which benchmarks six representative log anomaly detection methods proposed before 2016. Different from them, we focus on the latest deep learning-based approaches and investigate more practical issues such as unprecedented logs in testing data and inevitable anomalies in training data. More recently, Yang et al.~\cite{yang2021interview} presented an interview study of how developers use execution logs in embedded software engineering, summarizing the major challenges of log analysis. He et al.~\cite{he2021survey} conducted a comprehensive survey on log analysis for reliability engineering, covering the entire lifecycle of logs, including logging, log compression, log parsing, and various log mining tasks. Candido et al.~\cite{candido2019contemporary} provided a similar review for software monitoring. Other studies include cloud system attacks~\cite{khan2016cloud}, cyber security applications~\cite{landauer2020system}, etc.

%% file: secs/conclusion.tex
\section{Conclusion}
\label{sec:conclusion}

Logs have been widely used in various maintenance tasks for software systems. Due to their unprecedented volume, log-based anomaly detection on modern software systems is overwhelming the existing statistical and traditional machine learning-based methods. To pursue more intelligent solutions, many efforts have been devoted to developing deep learning-based anomaly detectors. However, we observe they are not fully deployed in industrial practices, which requires site reliability engineers to have a comprehensive knowledge of DL techniques. To fill this significant gap, in this paper, we conduct a detailed review of five popular neural networks for log-based anomaly detection and evaluate six state-of-the-art methods in terms of accuracy, robustness, and efficiency. Particularly, we explore whether logs' semantics can bring performance gain and whether they can help alleviate the issue of log instability. We also compare DL-based methods against their traditional ML-based counterparts. The results demonstrate that logs' semantics indeed improves models' robustness against noises in both training and testing data. Furthermore, we release an open-source toolkit of the studied methods to pave the way for model customization and improvement for both academy and industry.

%% file: acm-authordraft.bbl

\begin{thebibliography}{49}


\ifx \showCODEN    \undefined \def \showCODEN     #1{\unskip}     \fi
\ifx \showDOI      \undefined \def \showDOI       #1{#1}\fi
\ifx \showISBNx    \undefined \def \showISBNx     #1{\unskip}     \fi
\ifx \showISBNxiii \undefined \def \showISBNxiii  #1{\unskip}     \fi
\ifx \showISSN     \undefined \def \showISSN      #1{\unskip}     \fi
\ifx \showLCCN     \undefined \def \showLCCN      #1{\unskip}     \fi
\ifx \shownote     \undefined \def \shownote      #1{#1}          \fi
\ifx \showarticletitle \undefined \def \showarticletitle #1{#1}   \fi
\ifx \showURL      \undefined \def \showURL       {\relax}        \fi
\providecommand\bibfield[2]{#2}
\providecommand\bibinfo[2]{#2}
\providecommand\natexlab[1]{#1}
\providecommand\showeprint[2][]{arXiv:#2}

\bibitem[\protect\citeauthoryear{Bahdanau, Cho, and Bengio}{Bahdanau
  et~al\mbox{.}}{2014}]%
        {bahdanau2014neural}
\bibfield{author}{\bibinfo{person}{Dzmitry Bahdanau},
  \bibinfo{person}{Kyunghyun Cho}, {and} \bibinfo{person}{Yoshua Bengio}.}
  \bibinfo{year}{2014}\natexlab{}.
\newblock \showarticletitle{Neural machine translation by jointly learning to
  align and translate}.
\newblock \bibinfo{journal}{\emph{arXiv}} (\bibinfo{year}{2014}).
\newblock


\bibitem[\protect\citeauthoryear{Candido, Aniche, and van Deursen}{Candido
  et~al\mbox{.}}{2019}]%
        {candido2019contemporary}
\bibfield{author}{\bibinfo{person}{Jeanderson Candido},
  \bibinfo{person}{Maur{\'\i}cio Aniche}, {and} \bibinfo{person}{Arie van
  Deursen}.} \bibinfo{year}{2019}\natexlab{}.
\newblock \showarticletitle{Contemporary software monitoring: A systematic
  literature review}.
\newblock \bibinfo{journal}{\emph{arXiv}} (\bibinfo{year}{2019}).
\newblock


\bibitem[\protect\citeauthoryear{Chen, Kang, Li, Zhang, Zhang, Xu, Zhou, Yang,
  et~al\mbox{.}}{Chen et~al\mbox{.}}{2020}]%
        {chen2020towards}
\bibfield{author}{\bibinfo{person}{Zhuangbin Chen}, \bibinfo{person}{Yu Kang},
  \bibinfo{person}{Liqun Li}, \bibinfo{person}{Xu Zhang},
  \bibinfo{person}{Hongyu Zhang}, \bibinfo{person}{Hui Xu},
  \bibinfo{person}{Yangfan Zhou}, \bibinfo{person}{Li Yang}, {et~al\mbox{.}}}
  \bibinfo{year}{2020}\natexlab{}.
\newblock \showarticletitle{Towards intelligent incident management: why we
  need it and how we make it}. In \bibinfo{booktitle}{\emph{Proc. of
  ESEC/FSE'20}}. \bibinfo{pages}{1487--1497}.
\newblock


\bibitem[\protect\citeauthoryear{Dai, Li, Chen, Shang, and Chen}{Dai
  et~al\mbox{.}}{2020}]%
        {dai2020logram}
\bibfield{author}{\bibinfo{person}{Hetong Dai}, \bibinfo{person}{Heng Li},
  \bibinfo{person}{Che~Shao Chen}, \bibinfo{person}{Weiyi Shang}, {and}
  \bibinfo{person}{Tse-Hsun Chen}.} \bibinfo{year}{2020}\natexlab{}.
\newblock \showarticletitle{Logram: Efficient log parsing using n-gram
  dictionaries}.
\newblock \bibinfo{journal}{\emph{TSE}} (\bibinfo{year}{2020}).
\newblock


\bibitem[\protect\citeauthoryear{Dang, Lin, and Huang}{Dang
  et~al\mbox{.}}{2019}]%
        {dang2019aiops}
\bibfield{author}{\bibinfo{person}{Yingnong Dang}, \bibinfo{person}{Qingwei
  Lin}, {and} \bibinfo{person}{Peng Huang}.} \bibinfo{year}{2019}\natexlab{}.
\newblock \showarticletitle{AIOps: real-world challenges and research
  innovations}. In \bibinfo{booktitle}{\emph{Proc. of ICSE-C'19}}. IEEE,
  \bibinfo{pages}{4--5}.
\newblock


\bibitem[\protect\citeauthoryear{Du, Li, Zheng, and Srikumar}{Du
  et~al\mbox{.}}{2017}]%
        {du2017deeplog}
\bibfield{author}{\bibinfo{person}{Min Du}, \bibinfo{person}{Feifei Li},
  \bibinfo{person}{Guineng Zheng}, {and} \bibinfo{person}{Vivek Srikumar}.}
  \bibinfo{year}{2017}\natexlab{}.
\newblock \showarticletitle{Deeplog: Anomaly detection and diagnosis from
  system logs through deep learning}. In \bibinfo{booktitle}{\emph{Proc. of
  CCS'17}}. \bibinfo{pages}{1285--1298}.
\newblock


\bibitem[\protect\citeauthoryear{Farzad and Gulliver}{Farzad and
  Gulliver}{2020}]%
        {farzad2020unsupervised}
\bibfield{author}{\bibinfo{person}{Amir Farzad} {and} \bibinfo{person}{T~Aaron
  Gulliver}.} \bibinfo{year}{2020}\natexlab{}.
\newblock \showarticletitle{Unsupervised log message anomaly detection}.
\newblock \bibinfo{journal}{\emph{ICT Express}} \bibinfo{volume}{6},
  \bibinfo{number}{3} (\bibinfo{year}{2020}), \bibinfo{pages}{229--237}.
\newblock


\bibitem[\protect\citeauthoryear{Fu, Lou, Wang, and Li}{Fu
  et~al\mbox{.}}{2009}]%
        {fu2009execution}
\bibfield{author}{\bibinfo{person}{Qiang Fu}, \bibinfo{person}{Jian-Guang Lou},
  \bibinfo{person}{Yi Wang}, {and} \bibinfo{person}{Jiang Li}.}
  \bibinfo{year}{2009}\natexlab{}.
\newblock \showarticletitle{Execution anomaly detection in distributed systems
  through unstructured log analysis}. In \bibinfo{booktitle}{\emph{Proc. of
  ICDM'09}}. IEEE, \bibinfo{pages}{149--158}.
\newblock


\bibitem[\protect\citeauthoryear{Fu, Zhu, Hu, Lou, Ding, Lin, Zhang, and
  Xie}{Fu et~al\mbox{.}}{2014}]%
        {fu2014developers}
\bibfield{author}{\bibinfo{person}{Qiang Fu}, \bibinfo{person}{Jieming Zhu},
  \bibinfo{person}{Wenlu Hu}, \bibinfo{person}{Jian-Guang Lou},
  \bibinfo{person}{Rui Ding}, \bibinfo{person}{Qingwei Lin},
  \bibinfo{person}{Dongmei Zhang}, {and} \bibinfo{person}{Tao Xie}.}
  \bibinfo{year}{2014}\natexlab{}.
\newblock \showarticletitle{Where do developers log? an empirical study on
  logging practices in industry}. In \bibinfo{booktitle}{\emph{Proc. of
  ICSE-C'14}}. \bibinfo{pages}{24--33}.
\newblock


\bibitem[\protect\citeauthoryear{Hansen and Atkins}{Hansen and Atkins}{1993}]%
        {hansen1993automated}
\bibfield{author}{\bibinfo{person}{Stephen~E Hansen} {and}
  \bibinfo{person}{E~Todd Atkins}.} \bibinfo{year}{1993}\natexlab{}.
\newblock \showarticletitle{Automated System Monitoring and Notification with
  Swatch.}. In \bibinfo{booktitle}{\emph{Proc. of LISA'93}},
  Vol.~\bibinfo{volume}{93}. \bibinfo{pages}{145--152}.
\newblock


\bibitem[\protect\citeauthoryear{He, He, Chen, Yang, Su, and Lyu}{He
  et~al\mbox{.}}{2021}]%
        {he2021survey}
\bibfield{author}{\bibinfo{person}{Shilin He}, \bibinfo{person}{Pinjia He},
  \bibinfo{person}{Zhuangbin Chen}, \bibinfo{person}{Tianyi Yang},
  \bibinfo{person}{Yuxin Su}, {and} \bibinfo{person}{Michael~R Lyu}.}
  \bibinfo{year}{2021}\natexlab{}.
\newblock \showarticletitle{A survey on automated log analysis for reliability
  engineering}.
\newblock \bibinfo{journal}{\emph{CSUR}} \bibinfo{volume}{54},
  \bibinfo{number}{6} (\bibinfo{year}{2021}), \bibinfo{pages}{1--37}.
\newblock


\bibitem[\protect\citeauthoryear{He, Lin, Lou, Zhang, Lyu, and Zhang}{He
  et~al\mbox{.}}{2018}]%
        {he2018identifying}
\bibfield{author}{\bibinfo{person}{Shilin He}, \bibinfo{person}{Qingwei Lin},
  \bibinfo{person}{Jian-Guang Lou}, \bibinfo{person}{Hongyu Zhang},
  \bibinfo{person}{Michael~R Lyu}, {and} \bibinfo{person}{Dongmei Zhang}.}
  \bibinfo{year}{2018}\natexlab{}.
\newblock \showarticletitle{Identifying impactful service system problems via
  log analysis}. In \bibinfo{booktitle}{\emph{Proc. of ESEC/FSE'18}}.
  \bibinfo{pages}{60--70}.
\newblock


\bibitem[\protect\citeauthoryear{He, Zhu, He, and Lyu}{He
  et~al\mbox{.}}{2016}]%
        {he2016experience}
\bibfield{author}{\bibinfo{person}{Shilin He}, \bibinfo{person}{Jieming Zhu},
  \bibinfo{person}{Pinjia He}, {and} \bibinfo{person}{Michael~R Lyu}.}
  \bibinfo{year}{2016}\natexlab{}.
\newblock \showarticletitle{Experience report: System log analysis for anomaly
  detection}. In \bibinfo{booktitle}{\emph{Proc. of ISSRE'16}}. IEEE,
  \bibinfo{pages}{207--218}.
\newblock


\bibitem[\protect\citeauthoryear{He, Zhu, He, and Lyu}{He
  et~al\mbox{.}}{2020}]%
        {he2020loghub}
\bibfield{author}{\bibinfo{person}{Shilin He}, \bibinfo{person}{Jieming Zhu},
  \bibinfo{person}{Pinjia He}, {and} \bibinfo{person}{Michael~R Lyu}.}
  \bibinfo{year}{2020}\natexlab{}.
\newblock \showarticletitle{Loghub: a large collection of system log datasets
  towards automated log analytics}.
\newblock \bibinfo{journal}{\emph{arXiv}} (\bibinfo{year}{2020}).
\newblock


\bibitem[\protect\citeauthoryear{Joulin, Grave, Bojanowski, Douze, J{\'e}gou,
  and Mikolov}{Joulin et~al\mbox{.}}{2016}]%
        {joulin2016fasttext}
\bibfield{author}{\bibinfo{person}{Armand Joulin}, \bibinfo{person}{Edouard
  Grave}, \bibinfo{person}{Piotr Bojanowski}, \bibinfo{person}{Matthijs Douze},
  \bibinfo{person}{H{\'e}rve J{\'e}gou}, {and} \bibinfo{person}{Tomas
  Mikolov}.} \bibinfo{year}{2016}\natexlab{}.
\newblock \showarticletitle{Fasttext. zip: Compressing text classification
  models}.
\newblock \bibinfo{journal}{\emph{arXiv}} (\bibinfo{year}{2016}).
\newblock


\bibitem[\protect\citeauthoryear{Khan, Gani, Wahab, Bagiwa, Shiraz, Khan,
  Buyya, and Zomaya}{Khan et~al\mbox{.}}{2016}]%
        {khan2016cloud}
\bibfield{author}{\bibinfo{person}{Suleman Khan}, \bibinfo{person}{Abdullah
  Gani}, \bibinfo{person}{Ainuddin Wahid~Abdul Wahab},
  \bibinfo{person}{Mustapha~Aminu Bagiwa}, \bibinfo{person}{Muhammad Shiraz},
  \bibinfo{person}{Samee~U Khan}, \bibinfo{person}{Rajkumar Buyya}, {and}
  \bibinfo{person}{Albert~Y Zomaya}.} \bibinfo{year}{2016}\natexlab{}.
\newblock \showarticletitle{Cloud log forensics: foundations, state of the art,
  and future directions}.
\newblock \bibinfo{journal}{\emph{CSUR}} \bibinfo{volume}{49},
  \bibinfo{number}{1} (\bibinfo{year}{2016}), \bibinfo{pages}{1--42}.
\newblock


\bibitem[\protect\citeauthoryear{Landauer, Skopik, Wurzenberger, and
  Rauber}{Landauer et~al\mbox{.}}{2020}]%
        {landauer2020system}
\bibfield{author}{\bibinfo{person}{Max Landauer}, \bibinfo{person}{Florian
  Skopik}, \bibinfo{person}{Markus Wurzenberger}, {and}
  \bibinfo{person}{Andreas Rauber}.} \bibinfo{year}{2020}\natexlab{}.
\newblock \showarticletitle{System log clustering approaches for cyber security
  applications: A survey}.
\newblock \bibinfo{journal}{\emph{Computers \& Security}}  \bibinfo{volume}{92}
  (\bibinfo{year}{2020}), \bibinfo{pages}{101739}.
\newblock


\bibitem[\protect\citeauthoryear{LeCun, Bottou, Bengio, and Haffner}{LeCun
  et~al\mbox{.}}{1998}]%
        {lecun1998gradient}
\bibfield{author}{\bibinfo{person}{Yann LeCun}, \bibinfo{person}{L{\'e}on
  Bottou}, \bibinfo{person}{Yoshua Bengio}, {and} \bibinfo{person}{Patrick
  Haffner}.} \bibinfo{year}{1998}\natexlab{}.
\newblock \showarticletitle{Gradient-based learning applied to document
  recognition}.
\newblock \bibinfo{journal}{\emph{Proc. of the IEEE}} \bibinfo{volume}{86},
  \bibinfo{number}{11} (\bibinfo{year}{1998}), \bibinfo{pages}{2278--2324}.
\newblock


\bibitem[\protect\citeauthoryear{Liang, Zhang, Xiong, and Sahoo}{Liang
  et~al\mbox{.}}{2007}]%
        {liang2007failure}
\bibfield{author}{\bibinfo{person}{Yinglung Liang}, \bibinfo{person}{Yanyong
  Zhang}, \bibinfo{person}{Hui Xiong}, {and} \bibinfo{person}{Ramendra Sahoo}.}
  \bibinfo{year}{2007}\natexlab{}.
\newblock \showarticletitle{Failure prediction in ibm bluegene/l event logs}.
  In \bibinfo{booktitle}{\emph{Proc. of ICDM'07}}. IEEE,
  \bibinfo{pages}{583--588}.
\newblock


\bibitem[\protect\citeauthoryear{Lin, Zhang, Lou, Zhang, and Chen}{Lin
  et~al\mbox{.}}{2016}]%
        {lin2016log}
\bibfield{author}{\bibinfo{person}{Qingwei Lin}, \bibinfo{person}{Hongyu
  Zhang}, \bibinfo{person}{Jian-Guang Lou}, \bibinfo{person}{Yu Zhang}, {and}
  \bibinfo{person}{Xuewei Chen}.} \bibinfo{year}{2016}\natexlab{}.
\newblock \showarticletitle{Log clustering based problem identification for
  online service systems}. In \bibinfo{booktitle}{\emph{Proc. of ICSE-C'16}}.
  IEEE, \bibinfo{pages}{102--111}.
\newblock


\bibitem[\protect\citeauthoryear{Liu, Ting, and Zhou}{Liu
  et~al\mbox{.}}{2008}]%
        {liu2008isolation}
\bibfield{author}{\bibinfo{person}{Fei~Tony Liu}, \bibinfo{person}{Kai~Ming
  Ting}, {and} \bibinfo{person}{Zhi-Hua Zhou}.}
  \bibinfo{year}{2008}\natexlab{}.
\newblock \showarticletitle{Isolation forest}. In
  \bibinfo{booktitle}{\emph{Proc. of ICDM'08}}. IEEE,
  \bibinfo{pages}{413--422}.
\newblock


\bibitem[\protect\citeauthoryear{Lou, Fu, Yang, Xu, and Li}{Lou
  et~al\mbox{.}}{2010}]%
        {lou2010mining}
\bibfield{author}{\bibinfo{person}{Jian-Guang Lou}, \bibinfo{person}{Qiang Fu},
  \bibinfo{person}{Shengqi Yang}, \bibinfo{person}{Ye Xu}, {and}
  \bibinfo{person}{Jiang Li}.} \bibinfo{year}{2010}\natexlab{}.
\newblock \showarticletitle{Mining Invariants from Console Logs for System
  Problem Detection.}. In \bibinfo{booktitle}{\emph{Proc. of USENIXATC'10}}.
  \bibinfo{pages}{1--14}.
\newblock


\bibitem[\protect\citeauthoryear{Lu, Wei, Li, and Wang}{Lu
  et~al\mbox{.}}{2018}]%
        {lu2018detecting}
\bibfield{author}{\bibinfo{person}{Siyang Lu}, \bibinfo{person}{Xiang Wei},
  \bibinfo{person}{Yandong Li}, {and} \bibinfo{person}{Liqiang Wang}.}
  \bibinfo{year}{2018}\natexlab{}.
\newblock \showarticletitle{Detecting anomaly in big data system logs using
  convolutional neural network}. In \bibinfo{booktitle}{\emph{Proc. of
  DASC/PiCom/DataCom/CyberSciTech'18}}. IEEE, \bibinfo{pages}{151--158}.
\newblock


\bibitem[\protect\citeauthoryear{Meng, Liu, Zhu, Zhang, Pei, Liu, Chen,
  et~al\mbox{.}}{Meng et~al\mbox{.}}{2019}]%
        {meng2019loganomaly}
\bibfield{author}{\bibinfo{person}{Weibin Meng}, \bibinfo{person}{Ying Liu},
  \bibinfo{person}{Yichen Zhu}, \bibinfo{person}{Shenglin Zhang},
  \bibinfo{person}{Dan Pei}, \bibinfo{person}{Yuqing Liu},
  \bibinfo{person}{Yihao Chen}, {et~al\mbox{.}}}
  \bibinfo{year}{2019}\natexlab{}.
\newblock \showarticletitle{LogAnomaly: Unsupervised Detection of Sequential
  and Quantitative Anomalies in Unstructured Logs.}. In
  \bibinfo{booktitle}{\emph{Proc. of IJCAI'19}}. \bibinfo{pages}{4739--4745}.
\newblock


\bibitem[\protect\citeauthoryear{Nandi, Mandal, Atreja, Dasgupta, and
  Bhattacharya}{Nandi et~al\mbox{.}}{2016}]%
        {nandi2016anomaly}
\bibfield{author}{\bibinfo{person}{Animesh Nandi}, \bibinfo{person}{Atri
  Mandal}, \bibinfo{person}{Shubham Atreja}, \bibinfo{person}{Gargi~B
  Dasgupta}, {and} \bibinfo{person}{Subhrajit Bhattacharya}.}
  \bibinfo{year}{2016}\natexlab{}.
\newblock \showarticletitle{Anomaly detection using program control flow graph
  mining from execution logs}. In \bibinfo{booktitle}{\emph{Proc. of
  SIGKDD'16}}. \bibinfo{pages}{215--224}.
\newblock


\bibitem[\protect\citeauthoryear{Nedelkoski, Bogatinovski, Acker, Cardoso, and
  Kao}{Nedelkoski et~al\mbox{.}}{2020}]%
        {nedelkoski2020self}
\bibfield{author}{\bibinfo{person}{Sasho Nedelkoski}, \bibinfo{person}{Jasmin
  Bogatinovski}, \bibinfo{person}{Alexander Acker}, \bibinfo{person}{Jorge
  Cardoso}, {and} \bibinfo{person}{Odej Kao}.} \bibinfo{year}{2020}\natexlab{}.
\newblock \showarticletitle{Self-attentive classification-based anomaly
  detection in unstructured logs}.
\newblock \bibinfo{journal}{\emph{arXiv}} (\bibinfo{year}{2020}).
\newblock


\bibitem[\protect\citeauthoryear{Nguyen, Walde, and Vu}{Nguyen
  et~al\mbox{.}}{2016}]%
        {nguyen2016integrating}
\bibfield{author}{\bibinfo{person}{Kim~Anh Nguyen}, \bibinfo{person}{Sabine
  Schulte~im Walde}, {and} \bibinfo{person}{Ngoc~Thang Vu}.}
  \bibinfo{year}{2016}\natexlab{}.
\newblock \showarticletitle{Integrating distributional lexical contrast into
  word embeddings for antonym-synonym distinction}.
\newblock \bibinfo{journal}{\emph{arXiv}} (\bibinfo{year}{2016}).
\newblock


\bibitem[\protect\citeauthoryear{Oliner and Stearley}{Oliner and
  Stearley}{2007}]%
        {oliner2007supercomputers}
\bibfield{author}{\bibinfo{person}{Adam Oliner} {and} \bibinfo{person}{Jon
  Stearley}.} \bibinfo{year}{2007}\natexlab{}.
\newblock \showarticletitle{What supercomputers say: A study of five system
  logs}. In \bibinfo{booktitle}{\emph{Proc. of DSN'07}}. IEEE,
  \bibinfo{pages}{575--584}.
\newblock


\bibitem[\protect\citeauthoryear{Pennington, Socher, and Manning}{Pennington
  et~al\mbox{.}}{2014}]%
        {pennington2014glove}
\bibfield{author}{\bibinfo{person}{Jeffrey Pennington},
  \bibinfo{person}{Richard Socher}, {and} \bibinfo{person}{Christopher~D
  Manning}.} \bibinfo{year}{2014}\natexlab{}.
\newblock \showarticletitle{Glove: Global vectors for word representation}. In
  \bibinfo{booktitle}{\emph{Proc. of EMNLP'14}}. \bibinfo{pages}{1532--1543}.
\newblock


\bibitem[\protect\citeauthoryear{Prewett}{Prewett}{2003}]%
        {prewett2003analyzing}
\bibfield{author}{\bibinfo{person}{James~E Prewett}.}
  \bibinfo{year}{2003}\natexlab{}.
\newblock \showarticletitle{Analyzing cluster log files using logsurfer}. In
  \bibinfo{booktitle}{\emph{Proc. of the 4th Annual Conference on Linux
  Clusters}}. Citeseer.
\newblock


\bibitem[\protect\citeauthoryear{PyTorch}{PyTorch}{2016}]%
        {pytorch}
\bibfield{author}{\bibinfo{person}{PyTorch}.} \bibinfo{year}{2016}\natexlab{}.
\newblock \bibinfo{title}{[Online]}.
\newblock \bibinfo{howpublished}{\url{https://pytorch.org/}}.
\newblock


\bibitem[\protect\citeauthoryear{Rumelhart, Hinton, and Williams}{Rumelhart
  et~al\mbox{.}}{1985}]%
        {rumelhart1985learning}
\bibfield{author}{\bibinfo{person}{David~E Rumelhart},
  \bibinfo{person}{Geoffrey~E Hinton}, {and} \bibinfo{person}{Ronald~J
  Williams}.} \bibinfo{year}{1985}\natexlab{}.
\newblock \bibinfo{booktitle}{\emph{Learning internal representations by error
  propagation}}.
\newblock \bibinfo{type}{{T}echnical {R}eport}.
  \bibinfo{institution}{California Univ San Diego La Jolla Inst for Cognitive
  Science}.
\newblock


\bibitem[\protect\citeauthoryear{Russo, Succi, and Pedrycz}{Russo
  et~al\mbox{.}}{2015}]%
        {russo2015mining}
\bibfield{author}{\bibinfo{person}{Barbara Russo}, \bibinfo{person}{Giancarlo
  Succi}, {and} \bibinfo{person}{Witold Pedrycz}.}
  \bibinfo{year}{2015}\natexlab{}.
\newblock \showarticletitle{Mining system logs to learn error predictors: a
  case study of a telemetry system}.
\newblock \bibinfo{journal}{\emph{Empirical Software Engineering}}
  \bibinfo{volume}{20}, \bibinfo{number}{4} (\bibinfo{year}{2015}),
  \bibinfo{pages}{879--927}.
\newblock


\bibitem[\protect\citeauthoryear{Sahoo, Oliner, Rish, Gupta, Moreira, Ma,
  Vilalta, and Sivasubramaniam}{Sahoo et~al\mbox{.}}{2003}]%
        {sahoo2003critical}
\bibfield{author}{\bibinfo{person}{Ramendra~K Sahoo}, \bibinfo{person}{Adam~J
  Oliner}, \bibinfo{person}{Irina Rish}, \bibinfo{person}{Manish Gupta},
  \bibinfo{person}{Jos{\'e}~E Moreira}, \bibinfo{person}{Sheng Ma},
  \bibinfo{person}{Ricardo Vilalta}, {and} \bibinfo{person}{Anand
  Sivasubramaniam}.} \bibinfo{year}{2003}\natexlab{}.
\newblock \showarticletitle{Critical event prediction for proactive management
  in large-scale computer clusters}. In \bibinfo{booktitle}{\emph{Proc. of
  SIGKDD'03}}. \bibinfo{pages}{426--435}.
\newblock


\bibitem[\protect\citeauthoryear{Templates}{Templates}{2011}]%
        {message_templates}
\bibfield{author}{\bibinfo{person}{Message Templates}.}
  \bibinfo{year}{2011}\natexlab{}.
\newblock \bibinfo{title}{[Online]}.
\newblock \bibinfo{howpublished}{\url{https://messagetemplates.org/}}.
\newblock


\bibitem[\protect\citeauthoryear{Tensorflow}{Tensorflow}{2015}]%
        {tensorflow}
\bibfield{author}{\bibinfo{person}{Tensorflow}.}
  \bibinfo{year}{2015}\natexlab{}.
\newblock \bibinfo{title}{[Online]}.
\newblock \bibinfo{howpublished}{\url{https://www.tensorflow.org/}}.
\newblock


\bibitem[\protect\citeauthoryear{Vaswani, Shazeer, Parmar, Uszkoreit, Jones,
  Gomez, Kaiser, and Polosukhin}{Vaswani et~al\mbox{.}}{2017}]%
        {NIPS2017_3f5ee243}
\bibfield{author}{\bibinfo{person}{Ashish Vaswani}, \bibinfo{person}{Noam
  Shazeer}, \bibinfo{person}{Niki Parmar}, \bibinfo{person}{Jakob Uszkoreit},
  \bibinfo{person}{Llion Jones}, \bibinfo{person}{Aidan~N Gomez},
  \bibinfo{person}{\L~ukasz Kaiser}, {and} \bibinfo{person}{Illia Polosukhin}.}
  \bibinfo{year}{2017}\natexlab{}.
\newblock \showarticletitle{Attention is All you Need}. In
  \bibinfo{booktitle}{\emph{Proc. of NIPS'17}}.
\newblock


\bibitem[\protect\citeauthoryear{Xia, Bai, Yin, Li, and Xu}{Xia
  et~al\mbox{.}}{2021}]%
        {xia2021loggan}
\bibfield{author}{\bibinfo{person}{Bin Xia}, \bibinfo{person}{Yuxuan Bai},
  \bibinfo{person}{Junjie Yin}, \bibinfo{person}{Yun Li}, {and}
  \bibinfo{person}{Jian Xu}.} \bibinfo{year}{2021}\natexlab{}.
\newblock \showarticletitle{LogGAN: A log-level generative adversarial network
  for anomaly detection using permutation event modeling}.
\newblock \bibinfo{journal}{\emph{Information Systems Frontiers}}
  \bibinfo{volume}{23}, \bibinfo{number}{2} (\bibinfo{year}{2021}),
  \bibinfo{pages}{285--298}.
\newblock


\bibitem[\protect\citeauthoryear{Xu, Huang, Fox, Patterson, and Jordan}{Xu
  et~al\mbox{.}}{2009}]%
        {xu2009detecting}
\bibfield{author}{\bibinfo{person}{Wei Xu}, \bibinfo{person}{Ling Huang},
  \bibinfo{person}{Armando Fox}, \bibinfo{person}{David Patterson}, {and}
  \bibinfo{person}{Michael~I Jordan}.} \bibinfo{year}{2009}\natexlab{}.
\newblock \showarticletitle{Detecting large-scale system problems by mining
  console logs}. In \bibinfo{booktitle}{\emph{Proc. of SOSP'09}}.
  \bibinfo{pages}{117--132}.
\newblock


\bibitem[\protect\citeauthoryear{Yang, Chen, Wang, Wang, Jiang, Dong, and
  Zhang}{Yang et~al\mbox{.}}{2021a}]%
        {yang2021semi}
\bibfield{author}{\bibinfo{person}{Lin Yang}, \bibinfo{person}{Junjie Chen},
  \bibinfo{person}{Zan Wang}, \bibinfo{person}{Weijing Wang},
  \bibinfo{person}{Jiajun Jiang}, \bibinfo{person}{Xuyuan Dong}, {and}
  \bibinfo{person}{Wenbin Zhang}.} \bibinfo{year}{2021}\natexlab{a}.
\newblock \showarticletitle{Semi-supervised log-based anomaly detection via
  probabilistic label estimation}. In \bibinfo{booktitle}{\emph{Proc. of
  ICSE'21}}. IEEE, \bibinfo{pages}{1448--1460}.
\newblock


\bibitem[\protect\citeauthoryear{Yang, Schiffelers, and Lukkien}{Yang
  et~al\mbox{.}}{2021b}]%
        {yang2021interview}
\bibfield{author}{\bibinfo{person}{Nan Yang}, \bibinfo{person}{Ramon
  Schiffelers}, {and} \bibinfo{person}{Johan Lukkien}.}
  \bibinfo{year}{2021}\natexlab{b}.
\newblock \showarticletitle{An interview study of how developers use execution
  logs in embedded software engineering}. In \bibinfo{booktitle}{\emph{Proc. of
  ICSE-SEIP'21}}. IEEE, \bibinfo{pages}{61--70}.
\newblock


\bibitem[\protect\citeauthoryear{Yen, Moh, and Moh}{Yen et~al\mbox{.}}{2019}]%
        {yen2019causalconvlstm}
\bibfield{author}{\bibinfo{person}{Steven Yen}, \bibinfo{person}{Melody Moh},
  {and} \bibinfo{person}{Teng-Sheng Moh}.} \bibinfo{year}{2019}\natexlab{}.
\newblock \showarticletitle{CausalConvLSTM: Semi-supervised log anomaly
  detection through sequence modeling}. In \bibinfo{booktitle}{\emph{Proc. of
  ICMLA'19}}. IEEE, \bibinfo{pages}{1334--1341}.
\newblock


\bibitem[\protect\citeauthoryear{Yuan, Mai, Xiong, Tan, Zhou, and
  Pasupathy}{Yuan et~al\mbox{.}}{2010}]%
        {yuan2010sherlog}
\bibfield{author}{\bibinfo{person}{Ding Yuan}, \bibinfo{person}{Haohui Mai},
  \bibinfo{person}{Weiwei Xiong}, \bibinfo{person}{Lin Tan},
  \bibinfo{person}{Yuanyuan Zhou}, {and} \bibinfo{person}{Shankar Pasupathy}.}
  \bibinfo{year}{2010}\natexlab{}.
\newblock \showarticletitle{Sherlog: error diagnosis by connecting clues from
  run-time logs}. In \bibinfo{booktitle}{\emph{Proc. of ASPLOS'10}}.
  \bibinfo{pages}{143--154}.
\newblock


\bibitem[\protect\citeauthoryear{Yuan, Park, Huang, Liu, Lee, Tang, Zhou, and
  Savage}{Yuan et~al\mbox{.}}{2012}]%
        {yuan2012conservative}
\bibfield{author}{\bibinfo{person}{Ding Yuan}, \bibinfo{person}{Soyeon Park},
  \bibinfo{person}{Peng Huang}, \bibinfo{person}{Yang Liu},
  \bibinfo{person}{Michael~M Lee}, \bibinfo{person}{Xiaoming Tang},
  \bibinfo{person}{Yuanyuan Zhou}, {and} \bibinfo{person}{Stefan Savage}.}
  \bibinfo{year}{2012}\natexlab{}.
\newblock \showarticletitle{Be conservative: Enhancing failure diagnosis with
  proactive logging}. In \bibinfo{booktitle}{\emph{Proc. of OSDI'12}}.
  \bibinfo{pages}{293--306}.
\newblock


\bibitem[\protect\citeauthoryear{Zhang, Xu, Lin, Qiao, Zhang, Dang, Xie, Yang,
  Cheng, Li, et~al\mbox{.}}{Zhang et~al\mbox{.}}{2019}]%
        {zhang2019robust}
\bibfield{author}{\bibinfo{person}{Xu Zhang}, \bibinfo{person}{Yong Xu},
  \bibinfo{person}{Qingwei Lin}, \bibinfo{person}{Bo Qiao},
  \bibinfo{person}{Hongyu Zhang}, \bibinfo{person}{Yingnong Dang},
  \bibinfo{person}{Chunyu Xie}, \bibinfo{person}{Xinsheng Yang},
  \bibinfo{person}{Qian Cheng}, \bibinfo{person}{Ze Li}, {et~al\mbox{.}}}
  \bibinfo{year}{2019}\natexlab{}.
\newblock \showarticletitle{Robust log-based anomaly detection on unstable log
  data}. In \bibinfo{booktitle}{\emph{Proc. of ESEC/FSE'19}}.
  \bibinfo{pages}{807--817}.
\newblock


\bibitem[\protect\citeauthoryear{Zhao, Wang, Li, Peng, Wang, Pan, Wu, Feng,
  et~al\mbox{.}}{Zhao et~al\mbox{.}}{2021}]%
        {zhao2021empirical}
\bibfield{author}{\bibinfo{person}{Nengwen Zhao}, \bibinfo{person}{Honglin
  Wang}, \bibinfo{person}{Zeyan Li}, \bibinfo{person}{Xiao Peng},
  \bibinfo{person}{Gang Wang}, \bibinfo{person}{Zhu Pan}, \bibinfo{person}{Yong
  Wu}, \bibinfo{person}{Zhen Feng}, {et~al\mbox{.}}}
  \bibinfo{year}{2021}\natexlab{}.
\newblock \showarticletitle{An empirical investigation of practical log anomaly
  detection for online service systems}. In \bibinfo{booktitle}{\emph{Proc. of
  ESEC/FSE'21}}. \bibinfo{pages}{1404--1415}.
\newblock


\bibitem[\protect\citeauthoryear{Zhou, Peng, Xie, Sun, Ji, Liu, Xiang, and
  He}{Zhou et~al\mbox{.}}{2019}]%
        {zhou2019latent}
\bibfield{author}{\bibinfo{person}{Xiang Zhou}, \bibinfo{person}{Xin Peng},
  \bibinfo{person}{Tao Xie}, \bibinfo{person}{Jun Sun}, \bibinfo{person}{Chao
  Ji}, \bibinfo{person}{Dewei Liu}, \bibinfo{person}{Qilin Xiang}, {and}
  \bibinfo{person}{Chuan He}.} \bibinfo{year}{2019}\natexlab{}.
\newblock \showarticletitle{Latent error prediction and fault localization for
  microservice applications by learning from system trace logs}. In
  \bibinfo{booktitle}{\emph{Proc. of ESEC/FSE'19}}. \bibinfo{pages}{683--694}.
\newblock


\bibitem[\protect\citeauthoryear{Zhu, He, Fu, Zhang, Lyu, and Zhang}{Zhu
  et~al\mbox{.}}{2015}]%
        {zhu2015learning}
\bibfield{author}{\bibinfo{person}{Jieming Zhu}, \bibinfo{person}{Pinjia He},
  \bibinfo{person}{Qiang Fu}, \bibinfo{person}{Hongyu Zhang},
  \bibinfo{person}{Michael~R Lyu}, {and} \bibinfo{person}{Dongmei Zhang}.}
  \bibinfo{year}{2015}\natexlab{}.
\newblock \showarticletitle{Learning to log: Helping developers make informed
  logging decisions}. In \bibinfo{booktitle}{\emph{Proc. of ICSE'15}},
  Vol.~\bibinfo{volume}{1}. IEEE, \bibinfo{pages}{415--425}.
\newblock


\bibitem[\protect\citeauthoryear{Zhu, He, Liu, He, Xie, Zheng, and Lyu}{Zhu
  et~al\mbox{.}}{2019}]%
        {zhu2019tools}
\bibfield{author}{\bibinfo{person}{Jieming Zhu}, \bibinfo{person}{Shilin He},
  \bibinfo{person}{Jinyang Liu}, \bibinfo{person}{Pinjia He},
  \bibinfo{person}{Qi Xie}, \bibinfo{person}{Zibin Zheng}, {and}
  \bibinfo{person}{Michael~R Lyu}.} \bibinfo{year}{2019}\natexlab{}.
\newblock \showarticletitle{Tools and benchmarks for automated log parsing}. In
  \bibinfo{booktitle}{\emph{Proc. of ICSE-SEIP'19}}. IEEE,
  \bibinfo{pages}{121--130}.
\newblock


\end{thebibliography}
